\documentclass[twocolumn]{autart}    

\usepackage{nicefrac}
\usepackage{graphicx}
\usepackage{amssymb}
\usepackage{graphics} 
\usepackage{epsfig} 
\usepackage{times} 
\usepackage{amsmath} 
\usepackage{amsfonts}  
\usepackage{subfig}
\usepackage{epstopdf}
\usepackage{enumerate}
\usepackage{color}
\usepackage{mathtools}
\usepackage{algorithm}
\usepackage[noend]{algpseudocode}
\usepackage{float} 
\usepackage{framed} 
\usepackage{picins}

\newcommand{\dep}{\operatorname{dep}}

\newtheorem{definition}{Definition}
\newtheorem{corollary}{Corollary}
\newtheorem{remark}{Remark}
\newtheorem{proposition}{Proposition}
\newtheorem{theorem}{Theorem}
\newtheorem{lemma}{Lemma}

\begin{document}

\begin{frontmatter}

\title{Scalable Resetting Algorithms for Synchronization of Pulse-Coupled Oscillators over Rooted Directed Graphs\thanksref{footnoteinfo}} 
\vspace{-0.5cm}
\thanks[footnoteinfo]{Corresponding author: M. U. Javed. Part of the result was presented at the 58th IEEE Conference on Decision and Control, Nice, France, 2019. Research was supported by NSF ECCS-1809315 and CNS-1947613 and CU Boulder ASIRT Seed Grant.} 
\author{Muhammad U. Javed}\ead{muhammad.javed@colorado.edu}, \quad 
\author{Jorge I. Poveda}\ead{jorge.poveda@colorado.edu},    \quad           
\author{Xudong Chen}\ead{xudong.chen@colorado.edu}  

\address{Department of Electrical, Computer, and Energy Engineering, University of Colorado, Boulder, CO 80309}             


\begin{abstract}                          
We  study  the  problem  of  robust global synchronization  of  pulse-coupled  oscillators  (PCOs) over directed graphs. It is known that when the digraphs are strongly connected, global synchronization can be achieved by using a class of deterministic set-valued reset controllers \cite{sync_hybrid_poveda}. However, for large-scale networks, these algorithms are not scalable because some of their tuning parameters have upper bounds of the order $\mathcal{O}\left(\frac{1}{N}\right)$, where $N$ is the number of agents. This paper resolves this scalability issue by presenting several new results in the context of global synchronization of PCOs with more general network topologies using deterministic and stochastic hybrid dynamical systems. First, we establish that similar deterministic resetting algorithms can achieve robust, global, and fixed-time synchronization in any \emph{rooted acyclic digraph}. Moreover, in this case we show that the synchronization dynamics are now scalable as the tuning parameters of the algorithm are \emph{network independent}, i.e., of order $\mathcal{O}(1)$. However, the algorithms cannot be further extended to all rooted digraphs. We establish this new impossibility result by introducing a counter example with a particular rooted digraph for which global synchronization cannot be achieved, irrespective of the tuning of the reset rule. Nevertheless, we show that if the resetting algorithms are modified by accommodating an Erd{\"o}s-Ren{\'y}i type random graph model, then the resulting stochastic resetting dynamics will guarantee global synchronization almost surely for all \emph{rooted digraphs} and, moreover, the tunable parameters of the dynamics are network independent. 
Stability and robustness properties of the resetting algorithms are studied using the tools from set-valued hybrid dynamical systems. Numerical simulations are provided at the end of the paper for demonstration of the main results. 
\end{abstract}

\end{frontmatter}

\section{Introduction}
A network of PCOs consists of $N$ periodic dynamical systems, also called agents, sharing information via a communication directed graph (digraph). In most of the standard models of PCOs, e.g., \cite{Kannapan,cyclic_network,sync_hybrid_poveda,Anton17,pagliari}, each agent has an individual state $\tau_i\in\mathbb{R}$, which evolves according to the following constrained continuous-time dynamics:

\vspace{-0.7cm}
\begin{equation}\label{flows_agents}
\tau_i\in[0,1)\implies\dot{\tau}_i=\frac{1}{T},~~\forall~i\in\{1,2,\ldots,N\},
\end{equation}

\vspace{-0.4cm}
where $T>0$ is the period of the oscillator, and $[0,1)$ is a normalized unit interval. When the state of an agent $i$ finishes an oscillation, it sends a pulse to its out-neighbor agents $j$, and it proceeds to instantaneously \emph{reset} its individual state back to zero:

\vspace{-0.6cm}
\begin{equation}\label{resets_uncoupled}
\tau_i=1~\implies~\tau_i^+=0.
\end{equation}

\vspace{-0.3cm}
After receiving the pulse, each out-neighbor $j$ of agent $i$ instantaneously updates its own state $\tau_j$ using an individual phase update rule (PR) $\tau_j\mapsto\mathcal{P}_j(\tau_j)$, which usually has the following form:

\vspace{-0.8cm}
\begin{equation}\label{PUR}
\tau_j^+= \mathcal{P}_j(\tau_j)= \left\{\begin{array}{cl}
B(\tau_j), &\text{if}~~ \tau_j \in [0, r_j),\\
F(\tau_j), &\text{if}~~ \tau_j \in [r_j,1).\\
\end{array} \right.
\end{equation}
The mapping $\tau_j\mapsto B_j(\tau_j)$ is commonly referred to as the \emph{backward mapping}, and it decreases the value of $\tau_j$. The mapping $\tau_j\mapsto F_j(\tau_j)$ is referred to as the \emph{forward mapping}, and it increases $\tau_j$ \cite{Kannapan}. Whether an agent $j$ implements the mapping $B_j$ or the mapping $F_j$, depends on the position of $\tau_j$ with respect to the constant $r_j\in[0,1]$, which partitions in equation \eqref{PUR} the normalized unit interval of each agent. In this way, PCOs can be seen as multi-agent dynamical systems that combine the continuous-time dynamics \eqref{flows_agents} and the discrete-time dynamics \eqref{resets_uncoupled}-\eqref{PUR}. As a consequence, they are naturally modeled as networked multi-agent hybrid dynamical systems \cite{SurveyPovedaHDS}, and their convergence and stability properties are highly dependent on the structures of the mappings $\mathcal{P}_j$ and the partitions induced by the tunable parameters $r_j$. Given that equations \eqref{flows_agents}-\eqref{PUR} are quite general, PCOs can be used to model different biological systems, including Cardiac pacemakers \cite{peskin}, rhythmic flashing of fireflies \cite{buck}, electrical signals of neurons \cite{Neuron1,neuron2}, and biological oscillators \cite{Mirollo1990}, to name just a few. Networks of PCOs have also found several applications in engineering systems, such as cellular mobile radio \cite{cellular}, sensor networks \cite{wang,Doyle1,Doyle2}, and autonomous vehicles \cite{leonard}. More recently, PCOs have also been used to model and design clock synchronization and coordination feedback-based mechanisms for multi-agent sampled-data systems \cite{AHDS15}, \cite{sync_hybrid_poveda}. 

\vspace{-0.2cm}
A particular feature of PCOs is that their individual states are confined to evolve in the normalized interval $[0,1]$. By embedding the closed interval to the unit circle and identifying the two points $0$ and $1$ with each other, the network of PCOs can  be viewed as a multi-agent system evolving on the $N$-torus, where the state $\tau_i$ of the $i^{th}$ agent evolves in the unit circle flowing in counter-clockwise direction with frequency $1/T$. In this way, achieving global synchronization of PCOs can be cast as a global stabilization problem on a smooth compact manifold \cite{ClocksSensitivty}, \cite{sync_hybrid_poveda}. It is well-known that there is no smooth continuous-time state-feedback control law that can solve, in a robust way, such type of stabilization problems \cite{dorfler2014synchronization}, \cite{ClocksSensitivty}, \cite{SontagGoodReview}. This impossibility result has  motivated the development of several synchronization  algorithms that relax the global convergence requirement and, instead, focus on achieving only \emph{local} convergence \cite{Kuramoto,Friedmanglobal,SyncPhillip,Kannapan,DoyleTAC} or \emph{almost global} synchronization results, i.e., synchronization from all initial conditions except possibly from those in a set of measure zero \cite{Nishimura,Sepulchre09,Sepulchre12,Anton17}. However, for applications where measurement noise or external disturbances are unavoidable, almost global convergence results can be problematic given that they lead to non-zero measure sets from which synchronization cannot be achieved under arbitrarily small disturbances \cite{SontagGoodReview,Mayhew10Thesis}. In particular, such problematic non-zero measure sets can be quite large in multi-agent systems when the number of agents is large. Other results have achieved stochastic global synchronization on the unit circle under an all-to-all communication assumption \cite{HartmanBook}, by considering undirected (or, more precisely, bidirectional) graphs~\cite{Klinglmayr_2012} or strongly connected digraphs \cite{Klinglmayr}.  

\vspace{-0.2cm}
On the other hand, it has been shown that in certain cases it is possible to achieve \emph{robust global} synchronization in networks of PCOs by exploiting the underlying hybrid dynamics of the system, and by considering a set-valued regularization of the discontinuous PR \eqref{PUR}. This approach has been pursued in \cite{cyclic_network} for PCOs characterized by cycle digraphs, and in \cite{NunezSync} for PCOs with a global cue in the network. More recently, set-valued hybrid models have been investigated in \cite{AHDS15,sync_hybrid_poveda} using (deterministic) binary phase update rules (BPRs) that satisfy $\mathcal{P}_j:[0,1]\rightrightarrows\{0,1\}$, i.e., mappings that reset the position of each agent to a given point in the unit circle that identifies the beginning and the end of the interval $[0,1]$. By using this type of resetting rule, also called \emph{strong firing} \cite{Nishimuradiss}, it was shown in \cite{AHDS15,sync_hybrid_poveda}  that global and robust \emph{finite-time} synchronization of homogeneous PCOs can be achieved if the underlying information flow topology is characterized by a directed \emph{strongly connected} graph and if all the tuning parameters $r_j$ satisfy an upper bound of order $\mathcal{O}\left(\frac{1}{N}\right)$, see, e.g., \cite[Thm. 1]{sync_hybrid_poveda}. Thus, as the size of the network increases, the set of feasible parameters go to zero, resulting in a scalability issue that holds even for strongly connected digraphs. Besides the scalability issue, it has also remained an open question what type of directed graphs, other than the strongly connected ones, are necessary and/or sufficient for synchronization in PCOs with binary resetting rules.  

\vspace{-0.2cm}
In this paper we address both questions at a time. We characterize a class of digraphs, namely graphs that are rooted acyclic, for which \emph{robust}, \emph{global}, and \emph{fixed-time} synchronization can be achieved by using resetting algorithms with BPRs. In this case, the algorithms are scalable because the tunable parameters $r_j$ are independent of network size. This result further allows us to extend the resetting algorithm to a stochastic setting, where a sequence of {\em i.i.d.} Bernoulli random variables is used by each agent $i$ to decide whether or not to send the impulses to the out-neighbors $j$ after resetting its own state via equation \eqref{resets_uncoupled}. Interestingly, by injecting this randomness into the networked system, synchronization can be achieved almost surely for the entire class of rooted digraphs. Moreover, we show that with such digraphs it is in general impossible to achieve global synchronization using the deterministic resetting algorithm. We outline below the main contributions of the paper:

\begin{enumerate}

\vspace{-0.2cm}
%
\item 
We show in Proposition~\ref{lem:rooted_necessity} that having a rooted digraph is \emph{necessary} for achieving global synchronization of PCOs using deterministic BPRs. However, as shown in Proposition~\ref{prop:rooted_notsufficient}, this condition is not sufficient, which is true regardless of the choices of tunable parameters of the BPRs. Note that the gap between necessity and sufficiency makes our problem different from standard consensus dynamics in Euclidean spaces where having a rooted digraph is generally sufficient for global synchronization.  

%
\item 
We show that if the underlying digraph is rooted acyclic, then the deterministic resetting algorithm achieves global and fixed-time synchronization. Moreover, the tuning parameters of individual PCOs are network independent. The result is formulated as Theorem \ref{main_theorem1}, and extended in Corollary \ref{condensationcorollary} to quasi-acyclic digraphs. In each case, we provide a clear characterization of the convergence time in terms of the depth of the digraph.

%
\item We show that in the stochastic setting (with random triggers of pulses), the corresponding resetting algorithm can achieve global synchronization almost surely for {\em all} rooted digraphs. Moreover, we show that the probability of the network reaching synchronization converges exponentially fast to one.
The tuning parameters are again independent of the network size. 
The result is in contrast with the counter-example provided in Proposition~\ref{prop:rooted_notsufficient} for the deterministic setting.  
%
\end{enumerate}
\vspace{-0.2cm}
By the nature of the dynamics of the PCOs, we combine graph theoretic tools \cite{FB-LNS} and set-valued hybrid dynamical system's (HDS) theory \cite{teel} in order to analyze the qualitative properties of the network. This formalism is instrumental in the robustness analysis of the synchronization dynamics with respect to small additive bounded disturbances that are unavoidable in practice. Such type of robustness results are critical for the safe implementation of the algorithms in practical applications.
To the best of author's knowledge, the results of this paper are the first ones that address the \emph{scalability} issue that emerges in the global synchronization problem of PCOs, and that establish \emph{robust global} synchronization over quasi-acyclic  digraphs without using leading agents or global cues, with an explicit  characterization of the convergence time as a function of the structure of the digraph. Moreover, unlike existing almost sure convergence results in the literature of stochastic synchronization of PCOs, e.g., \cite{Klinglmayr_2012,Klinglmayr,pagliari}, we use the framework of stochastic hybrid dynamical systems (SHDS) to establish uniform global asymptotic stability in probability of the PCOs with respect to the synchronization set. This property not only implies global synchronization almost surely, but also \emph{uniform stability in probability} in the synchronization set in the Lyapunov and Lagrange sense \cite{Teel2014StabilitySurvey}. Furthermore, 
we provide theoretical bounds for the stochastic synchronization time of our algorithms, as functions of the structures of the underlying digraphs.  Our preliminary results, reported in the conference paper \cite{umarCDC}, considered only deterministic algorithms and presented results only for rooted acyclic digraphs, a subclass of the digraphs considered in this paper. Analysis and proofs of the results were also omitted in the conference version.  

The rest of this paper is organized as follows: Section \ref{sec_preliminaries} presents some preliminaries. Main results for the deterministic and stochastic settings are presented and established in Sections \ref{section_deterministic_sync} and  \ref{section_stochastic_sync}, respectively. 
Section \ref{sec_examples} shows numerical examples. The paper ends with conclusions.  

{\bf Notations.} 
Given a vector $x$ in $\mathbb{R}^n$,  we let $|x|$ be the standard Euclidean norm of $x$.  
For a compact set $\mathcal{A} \subset \mathbb{R}^n$, we let $|x|_\mathcal{A} := \min_{y \in \mathcal{A}}|x -y|$.  We will also use $|\cdot|$ to denote the cardinality of a finite set. 
For constant vectors (e.g., parameters) we will use $c_n\in\mathbb{R}^n$ to denote a constant vector with all entries equal to $c\in\mathbb{R}$. We use $\mathbb{S}\subset\mathbb{R}^2$ to denote the unit circle centered at the origin, i.e., $\mathbb{S}:=\left\{(x_1,x_2)\in\mathbb{R}^2:x_1^2+x_2^2=1\right\}$. Given a set $B$, we use $B^N$ to denote the $N$-Cartesian product of $B$, i.e., $B^N=B\times B\times\cdots\times B$. A function $\beta$ is said to be of class $\mathcal{K}\mathcal{L}$ if it is non-decreasing in its first argument, non-increasing in its second argument, $\lim_{r\to0^+}\beta(r,s)=0$ for each $s\in\mathbb{R}_{\geq0}$, and $\lim_{s\to\infty}\beta(r,s)=0$ for each $r\in\mathbb{R}_{\geq0}$. For a closed set $B\subset\mathbb{R}^n$, and $\varepsilon>0$, $B+\varepsilon\mathbb{B}$ denotes the set $\{x\in\mathbb{R}^n:|x|_B\leq \varepsilon\}$.

\vspace{-0.2cm}
\section{Preliminaries}
\label{sec_preliminaries}

\vspace{-0.2cm}
This section presents basic notions from graph theory, preliminaries about deterministic and stochastic hybrid dynamical systems, and notions of system stability. 

\vspace{-0.2cm}
\subsection{Basic Notions from Graph Theory}
\vspace{-0.2cm}
 A directed graph, or digraph, is denoted by $\mathcal{G}:=(\mathcal{V},\mathcal{E})$, and it is characterized by the set of vertices $\mathcal{V}:=\{1,2,\ldots,N\}$, and the set of edges $\mathcal{E} \subset \mathcal{V} \times \mathcal{V}$. 
 In this paper, we consider only simple digraphs, i.e., digraphs without self-arcs. We adopt the convention that information flows from vertex $i$ to vertex $j$ if $(i,j) \in \mathcal{E}$, and we define $i$ as an in-neighbor of $j$, and $j$ is an out-neighbor of~$i$. 
 
 A \textit{walk} from a vertex $i$ to a vertex $j$, denoted by $w_{ij}$, is a sequence $\{{i_0},{i_1},\ldots, {i_m}\}$, with $i_0=i$ and $i_m=j$, in which each pair $(i_k,i_{k+1})\in\mathcal{E}$ for all $k \in \{0, 1,\cdots, m-1\}$. A \emph{path} corresponds to a walk in which all the vertices are pairwise distinct. A \textit{cycle} is a walk in which there is no repetition of vertices other than the repetition of the starting and ending vertex.  The \textit{length} of a path/cycle/walk is defined to be the number of edges in that path/cycle/walk. A vertex $i\in\mathcal{V}$ is said to be a {\em root} of $\mathcal{G}$ if for any other vertex $j\in\mathcal{V}$, there exists a path from $i$ to $j$. A digraph $\mathcal{G}$ with at least one root is a {\it rooted digraph}. A rooted digraph $\mathcal{G}$ without a cycle is \textit{rooted acyclic}. 
 If $\mathcal{G}$ is rooted acyclic, then there is a unique root. 
 
 In general, a rooted digraph $\mathcal{G}$ can have multiple roots. All the roots then form a strongly connected subgraph $\mathcal{G}_R$. 
We call $\mathcal{G}_R$ the {\em root component} of $\mathcal{G}$. The digraph $\mathcal{G}$ is said to be \textit{quasi-acyclic} if all the cycles of  $\mathcal{G}$ are contained in the root component. In other words, if we condense $\mathcal{G}_R$ into a single vertex, then the resulting condensed digraph, denoted by $\mathcal{G}_c$, is rooted acyclic.

A rooted acyclic digraph is a {\em directed tree} if every vertex, except the root, has a single in-neighbor. 
Every rooted digraph $\mathcal{G} = (\mathcal{V}, \mathcal{E})$ contains a directed tree $\mathcal{T} = (\mathcal{V}, \mathcal{E}')$, with the same vertex set, as its subgraph. We call $\mathcal{T}$ a {\em directed spanning tree}. 
Let $\mathcal{G}$ be a directed tree with $i^*$ the root. The \textit{depth} of a vertex $i$ other than $i^*$, denoted by $\dep(i)$, is the length of the unique path from $i^* $ to $i$. The depth of $i^*$ is $0$ by default. The \textit{depth of $\mathcal{G}$} is $\dep(\mathcal{G}) := \max_{v_i\in \mathcal{V}} \dep(i)  $. For the given directed tree $\mathcal{G}$, we decompose the vertex set $\mathcal{V}$  as $\mathcal{V} = \cup_{l=0}^{\dep(\mathcal{G})}\mathcal{V}_l$, where $\mathcal{V}_l$ contains all the vertices of depth~$l$. Let $\mathcal{G}$ be a rooted digraph. We define the depth of $\mathcal{G}$, denoted by $\dep(\mathcal{G})$, to be the maximal depth of a directed spanning tree $\mathcal{T}$ of $\mathcal{G}$. 



\vspace{-0.2cm}
\subsection{Hybrid Dynamical Systems with Random Inputs}
%

\vspace{-0.2cm}
A stochastic hybrid dynamical systems (SHDS) with state $x\in\mathbb{R}^n$ and random input $v\in\mathbb{R}^m$ is characterized by the following set of equations:
\begin{subequations}\label{SHDS1}
	\begin{align}
	&x\in C,~~~~~~~~~\dot{x}= f(x),\label{SHDS_flows0}\\
	&x\in D,~~~~~~x^+\in G(x,v^+),~~~v\sim \mu(\cdot)~~\label{SHDS_jumps0}
	\end{align}
\end{subequations}
where the function $f:\mathbb{R}^n\to\mathbb{R}^n$, called the \emph{flow map}, describes the continuous-time dynamics of the system; the set $C\subset\mathbb{R}^n$, called the \emph{flow set}, describes the points in the space where $x$ is allowed to evolve according to the differential equation \eqref{SHDS_flows0}; $G:\mathbb{R}^n\times\mathbb{R}^m\rightrightarrows\mathbb{R}^n$, called the jump map, is a set-valued mapping that characterizes the discrete-time dynamics of the system; and $D\subset\mathbb{R}^n$, called the jump set, describes the points in the space where $x$ is allowed to evolve according to the stochastic difference inclusion \eqref{SHDS_jumps0}. We use $v^+$ as a place holder for a sequence of independent, identically distributed ({\em i.i.d.}) input random variables $\{{\bf v_k}\}_{k=1}^{\infty}$ with probability distribution $\mu$, derived from an abstract probability space $(\Omega,\mathcal{F},\mathbb{P})$.  
General SHDS of the form \eqref{SHDS1} have been introduced and analyzed in \cite{rec_principle}.  In this paper we restrict our attention to SHDS that satisfy the following basic conditions:

\vspace{-0.2cm}
\begin{definition}[Basic Conditions]\label{definitionbasic1}
A SHDS is said to satisfy the \textbf{\textit{basic conditions}} if the following holds: (a) The sets $C$ and $D$ are closed,  $C\subset \text{dom}(f)$, and $D\subset\text{dom}(G)$. (b) The function $f$ is continuous.(c) The set-valued mapping $G:\mathbb{R}^n\times\mathbb{R}^m\rightrightarrows\mathbb{R}^n$ is locally bounded and the mapping $v\mapsto \text{graph}(G(\cdot,v)):=\{(x,y)\in\mathbb{R}^n\times\mathbb{R}^n:y\in G(x,v)\}$ is measurable with closed values.
\end{definition}
\vspace{-0.2cm}
When the discrete-time dynamics \eqref{SHDS_jumps0} do not depend on random inputs, the SHDS \eqref{SHDS1} is reduced to a standard deterministic HDS \cite{teel}:

\vspace{-1cm}
\begin{subequations}\label{HDS}
	\begin{align}
	&x\in C,~~~~~~~~~\dot{x}= f(x),\label{HDS_flows1}\\
	&x\in D,~~~~~~x^+\in G(x).~~\label{HDS_jumps1}
	\end{align}
\end{subequations}

\vspace{-0.5cm}
Solutions to hybrid systems (either stochastic \eqref{SHDS1} or deterministic \eqref{HDS}) are parameterized by both continuous- and discrete-time indices $t\in\mathbb{R}_{\geq0}$ and $j\in\mathbb{Z}_{\geq0}$. The index $t$ increases continuously during flows \eqref{HDS_flows1} or \eqref{SHDS_flows0}, and the index $j$ increases by one when a jump occurs via \eqref{HDS_jumps1} or \eqref{SHDS_jumps0}. 

Of particular interest to us are solutions that  have an unbounded time domain in both $t$ and $j$ directions.  Such type of solution is maximal (i.e., its domain is not a proper subset of the domain of other solution) and non-Zeno (i.e., they do not have accumulation points in $t$). For a precise definition of  maximal non-Zeno solutions $x$ to HDS of the form \eqref{HDS} we refer the reader to the Appendix \ref{solutionsHDS}. Similarly, for a precise definition of maximal random solutions $\mathbf{x}_{\omega}$ to SHDS of the form \eqref{SHDS1} we refer the reader to the Appendix \ref{solutions_SHDS}. 
\vspace{-0.2cm}
%
\subsection{Stability and Convergence Notions}

\vspace{-0.2cm}
In this paper, we will use the following standard stability notion~\cite{teel} for deterministic HDS \eqref{HDS}:
\begin{definition}\label{UGAS2}
A HDS $\mathcal{H}:=\{C,f,D,G\}$ is said to render a compact set $\mathcal{A}$ \textbf{uniformly globally asymptotically stable (UGAS)} if there exists a function $\beta\in\mathcal{KL}$ such that every solution $x$ of \eqref{HDS} satisfies the bound  
\begin{equation*}
|x(t,j)|_{\mathcal{A}}\leq \beta(|x(0,0)|_{\mathcal{A}},t+j),
\end{equation*}
for all $(t,j)\in\text{dom}(x)$. We say that $\mathcal{H}$ renders $\mathcal{A}$ \textbf{uniformly globally fixed-time stable (UGFxTS)} if, additionally, there exists a $\overline{T}>0$ such that $\beta(|x(0,0)|_{\mathcal{A}},t+j)=0$ for all $t+j\geq \overline{T}$ and all $x(0,0)\in C\cup D$.
\end{definition}
The UGAS stability property introduced in Definition \ref{UGAS2} is standard in the analysis of hybrid dynamical systems, see \cite[Chp.3]{teel}. On the other hand, the notion of UGFxTS is stronger, since it asks that every solution of the system should converge in finite time to the set $\mathcal{A}$, with a convergence time that can be upper bounded by a constant that is independent of the initial conditions. Note that when $C$ and $D$ are compact sets, fixed-time stability is equivalent to finite-time stability \cite{WangFiniteSync}. 


%

\vspace{-0.2cm}
To study the stability properties of SHDS of the form \eqref{SHDS1}, we use the following definition borrowed from \cite{sta_rec}: 

\vspace{-0.2cm}
\begin{definition}\label{SHDS2def}
A SHDS \eqref{SHDS1} is said to render a compact set $\mathcal{A}$:

\vspace{-0.1cm}
\begin{enumerate}[(a)]
\item \textbf{Uniformly Lyapunov stable in probability}  if for each $\varepsilon>0$ and $\rho>0$ there exists a $\delta>0$ such that for all $x_\omega(0,0) \in \mathcal{A}+\delta \mathbb{B}$, every maximal random solution ${\bf x}_\omega$ from $x_\omega(0,0)$ satisfies the inequality:
\begin{align}\label{UGASpDef}
&\mathbb{P}\Big(\mathbf{x}_{\omega}(t,j)\in\mathcal{A}+\varepsilon\mathbb{B}^\circ,~\forall~ (t,j)\in\emph{\text{dom}}(\mathbf{x}_{\omega})\Big)\notag\\
&~~~~~~~~~~~~~~~~~~~~~~~\geq 1-\rho.
\end{align}
\item \textbf{Uniformly Lagrange stable in probability} if for each $\delta>0$ and $\rho >0$, there exists $\varepsilon>0$ such that the inequality \eqref{UGASpDef} holds. 
\item \textbf{Uniformly globally attractive in probability} if for each $\varepsilon>0, \rho>0$ and $R>0$, there exists $\gamma\geq 0$ such that for all random solutions  ${\bf x}_\omega$ with $x_\omega(0,0)\in \mathcal{A}+R \mathbb{B}$ the following holds: 
\begin{align*}
\mathbb{P}\Big(&\mathbf{x}_{\omega}(t,j)\in\mathcal{A}+\varepsilon\mathbb{B}^\circ,\forall~t+j\geq \gamma,(t,j)\in \emph{\text{dom}}(\mathbf{x}_{\omega})\Big)\notag \\
&~~~~~~~~~~~~~~~~~~~~~~~\geq 1-\rho.
\end{align*}
\end{enumerate}
System \eqref{SHDS1} is said to render a compact set $\mathcal{A}\subset\mathbb{R}^n$ \textbf{Uniformly Globally Asymptotically Stable in Probability (\textbf{UGASp})} if it satisfies conditions (a), (b), and (c).
\end{definition}
Definition \ref{SHDS2def} is a natural extension of Definition \ref{UGAS2} to the stochastic domain. Moreover, under the Basic Conditions and certain causality assumptions on the solutions of the system, UGASp is a property that can be established by combining suitable Lyapunov functions and stochastic hybrid invariance principles \cite[Thm. 8]{rec_principle}. These tools will be instrumental in the analysis of our algorithms in the next sections.

\vspace{-0.2cm}
\section{Deterministic Resetting Algorithms}
\label{section_deterministic_sync}
\vspace{-0.2cm}
In this section we study how to construct deterministic BPRs that are scalable and that achieve robust global synchronization in PCOs over sparse networks.
\vspace{-0.2cm}
\subsection{Well-Posed Model for Robust Synchronization}
\vspace{-0.2cm}
We start by constructing suitable regularizations of discontinuous BPRs of the form \eqref{PUR}. First, we recall that if all the agents are completely decoupled, then their dynamics are described by~\eqref{flows_agents} and~\eqref{resets_uncoupled}. When agents are coupled through a network, every agent $i$ will send a pulse to its out-neighbors whenever $\tau_i$ reaches $1$ and resets the value to $0$ via \eqref{resets_uncoupled}. On the other hand, if agent $j$ receives a pulse from its  in-neighbor, then we assign the following set-valued BPR to the agent $j$:
\vspace{-0.2cm}
\begin{equation}\label{update_neigbors}
\tau_j^+\in \overline{\mathcal{P}}_j(\tau_j):= \left\{\begin{array}{cl}
\{0\} &~~ \tau_j \in [0, r_j)\\
\{0,1\} &~~ \tau_j=r_j\\
\{1\} &~~ \tau_j \in (r_j,1]\\
\end{array} \right.,
\end{equation}
where $r_j\in[0,1]$ is the tuning parameter. Each $r_j$ partitions $[0,1]$ into two segments. 
For convenience, we call  $r := [r_1;\ldots ; r_N]$ a \textit{\textbf{partition vector}}.  

\begin{remark}{\em 
As in \cite{AHDS15}, the set-valued mapping $\overline{\mathcal{P}}_j:[0,1]\rightrightarrows\{0,1\}$ in equation \eqref{update_neigbors} is generated as the outer semicontinuous hull\footnote{See Appendix A for a precise definition of the outer semicontinuous hull of a mapping.}of the BPR \eqref{PUR} with forward map $F_j(\tau_j)=1$ and backward map $B_j(\tau_j)=0$. This is a type of regularization used in the robustness analysis of discontinuous discrete-time dynamical systems \cite{Kellet}, which will allows to establish suitable robustness results for synchronization dynamics. Since in the unit circle the points $0$ and $1$ are identified to be the same, the mapping \eqref{update_neigbors} can be seen as a simple forward/backward resetting rule. 
}
\end{remark}
\vspace{-0.2cm}
We now model the dynamics of the agents, together with the BPRs \eqref{update_neigbors}, as a HDS of the form \eqref{HDS} with overall state $\tau=[\tau_1,\tau_2,\ldots,\tau_N]^\top$. Given a digraph $\mathcal{G}$ of $N$ vertices, and a tuning vector $r\in [0,1]^N$, we write the HDS as
\begin{equation}\label{eq:DHDS}
\mathcal{H}(r, \mathcal{G}):=\{C,f,D,G\},
\end{equation}
with flow and jump sets given by
\begin{align}\label{sets_deterministic}
C:=[0,1]^N,~\text{and}~D:=\big\{\tau\in C:\text{max}_{i\in\mathcal{V}}~\tau_i=1\big\},
\end{align}
respectively; flow map and jump map given by
\vspace{-0.3cm}
\begin{equation}\label{flow_map_deterministic}
f(\tau):=\frac{1}{T}\cdot \mathbf{1}_N,~~~~~G(\tau):=\overline{G^0(\tau)},
\end{equation}

\vspace{-0.7cm}
respectively, where $\overline{G^0}$ is the outer-semicontinuous hull of the set-valued mapping $G^0:[0,1]^N\rightrightarrows\mathbb{R}^N$ given by
\vspace{-0.2cm}
\begin{align}\label{initial_stochastic_jump_map}
G^0(\tau):=&\Bigg\{g\in\mathbb{R}^N:g_i=0,\notag\\
&~~g_j\in \left\{\begin{array}{ll} \overline{\mathcal{P}}_j(\tau_j), & (i,j)\in\mathcal{E}\\
\{\tau_j\}, &  (i,j)\notin\mathcal{E}
\end{array}\right\},~\forall~j\neq i\Bigg\},
\end{align}
which is defined to be nonempty only when $\tau_i=1$ for some $i\in\mathcal{V}$ and $\tau_j\in[0,1)$ for $j\neq i$. Importantly, as in \cite{sync_hybrid_poveda}, by construction of the jump set and the jump map, when more than two agents satisfy the condition $\tau_i=1$, their jumps will occur sequentially rather than in parallel. This behavior is induced on purpose to guarantee a suitable semi-continuous dependence on the initial conditions for the solutions of the system. 
Indeed, in order to capture the effect of arbitrarily small disturbances acting on the states of the PCOs, the synchronization model must guarantee that for each $\tau_0\in[0,1]$, and each graphically convergent sequence of solutions $\{\tau_k\}_{k\in\mathbb{Z}_{\geq0}}$ with components $\tau_{i,k}$ satisfying
\begin{equation}\label{seq_initial}
0\leq \tau_{i,k}(0,0)\leq \tau_{i+1,k}(0,0)\leq \ldots \leq \tau_{i+I,k}(0,0)<\tau_0,
\end{equation} 
for some $i\in\mathcal{V}$ and $I\in \mathbb{Z}_{>0}$, and
\begin{align*}
\lim_{k\to\infty}\tau_{i,k}(0,0)=&\lim_{k\to\infty}\tau_{i+1,k}(0,0)=\ldots\notag\\
&~~~~~~\ldots=\lim_{k\to\infty}\tau_{i+I,k}(0,0)=\tau_0,
\end{align*}
the sequence of solutions must graphically converge to a solution of the system starting from the set of initial conditions

\vspace{-0.7cm}
\begin{equation}\label{limiting_initial_condition}
\tau_i=\tau_{i+1}=\ldots=\tau_I=\tau_0.
\end{equation}

\vspace{-0.3cm}
This condition is particularly relevant for the case when $\tau_0=1$, since it implies that the states $\tau_{i,k}$ with initial conditions satisfying \eqref{seq_initial} are sequentially reset with smaller and smaller times between resets as $k\to\infty$. Thus, in the limiting case \eqref{limiting_initial_condition}, all agents must  reset their states but the resets must occur sequentially. Since no order is specified for the sequential jumps, and there is no reason to give priority to one agent over the other, any robust model of PCOs must take into account all the possible trajectories induced by the $N!$ different resetting orders that can emerge from condition \eqref{limiting_initial_condition} with $\tau_0=1$, $i=1$, and $I=N$. 
\begin{remark}\label{multiplesolutions}\emph{
The construction of $G$ suggests that studying the individual behavior of every possible solution of the system becomes intractable as $N$ increases. In order to address this issue, in this paper we will use Lyapunov stability theory to analyze the qualitative behavior of \emph{every} possible solution of the system} from any initial condition in $[0,1]^N$.
\end{remark}

The following fact follows directly from the construction:
\vspace{-0.2cm}
\begin{lemma}\label{well_posedHDS}
For every tuning vector $r$ and every digraph $\mathcal{G}$, the HDS $\mathcal{H}(r,\mathcal{G})$ introduced in~\eqref{eq:DHDS} satisfies the basic conditions.
\end{lemma}
\vspace{-0.2cm}
We aim to characterize pairs $(r, \mathcal{G})$ that make the corresponding HDS $\mathcal{H}(r, \mathcal{G})$ well behaved (non-Zeno behavior) and, moreover, render the following compact set UGAS: 
\vspace{-0.6cm}

\begin{equation}\label{synchronization_set}
\mathcal{A}_s:=\left([0,1]\cdot 1_N\right)\cup \{0,1\}^N.
\end{equation}
\vspace{-0.6cm}

It should be clear that if the state $\tau$ belongs to $\mathcal{A}_s$, then the network reaches synchronization. 
For convenience, we introduce the following definition:
\vspace{-0.2cm}
\begin{definition}\label{compatibility}
Let $\mathcal{A}_s$ be the compact set in~\eqref{synchronization_set}. Let $r\in [0,1]^N$ be a tuning vector and $\mathcal{G}$ be a digraph of $N$ vertices. The pair $(r, \mathcal{G})$ is a \textbf{sync-pair} if 
\begin{enumerate}[(a)]
\item For every initial condition in $C\cup D$ there exists a solution to $\mathcal{H}(r,\mathcal{G})$, and each solution has an unbounded time domain and it is uniformly non-Zeno;
\item The HDS $\mathcal{H}(r,\mathcal{G})$ renders $\mathcal{A}_s$ UGAS.
\end{enumerate}
\end{definition}

While the stability analysis of the set $\mathcal{A}_s$ will be highly dependent on the communication digraph $\mathcal{G}$ and the tuning vector $r$, the following Lemma will be instrumental in satisfaction of item (a) in Definition \ref{compatibility}. 
\vspace{-0.2cm}
\begin{lemma}\label{nonzeno}
Consider the HDS $\mathcal{H}(r,\mathcal{G})$. For any tuning vector $r\in(0,1]^N$ and any digraph $\mathcal{G}$ we have that item (a) in Definition \ref{compatibility} holds, and the number of jumps in any interval of length $T$ is bounded above by $N\underline{r}^{-1}$, where $\underline{r}:=\min_{i\in\mathcal{V}}r_i$. 
\end{lemma}
\textbf{Proof}: Let $\mathcal{G}$ be given, and let $r\in(0,1]^N$. Since the HDS $\mathcal{H}(r,\mathcal{G})$ satisfies the basic conditions, and since $f(\tau)>0$ for all $\tau\in[0,1)^N$, by \cite[Prop. 6.10]{teel} there exists at least one non-trivial solution from every initial condition $\tau(0,0)\in C\cup D$. Since the flow map is globally Lipschitz and the flow set is compact, every solution $\tau$ does not exhibit finite escape times. Moreover, since $G(D)\subset C\cup D$ solutions cannot stop due to jumps. Thus, by \cite[Prop. 6.10]{teel} every solution $\tau$ is complete, i.e., it has an unbounded time domain. To show absence of Zeno-behavior, it suffices to rule out the existence of discrete solutions to system $\mathcal{H}(r,\mathcal{G})$ \cite[Prop. 6.35]{teel}. Suppose by contradiction that there exists a maximal solution $\tau$ satisfying $\tau(0,j)\in D$ for all $j\in\mathbb{Z}_{\geq0}$. This implies that for all $j\in\mathbb{Z}_{\geq0}$ there exists some $i^*\in\mathcal{V}$ such that $\tau_{i^*}(0,j)=1$. By construction of the dynamics, and without loss of generality, we will have that $\tau_{i^*}(0,j+1)=0$, which implies that agent $i^*$ cannot trigger further jumps. This argument can be repeated at most $N-1$ consecutive times, after which all agents would satisfy $\tau_i\neq 1$, i.e., $\tau(0,j+N)\notin D$, which contradicts the original assumption. 

\vspace{-0.2cm}
To show the bound on the jumps, it suffices to observe that the fastest every agent can be in the jump set after resetting its own clock is by flowing $r_iT$ seconds to satisfy the condition $\tau\geq r_i$. Therefore, for any digraph $\mathcal{G}$ the maximum number of jumps that can occur in any interval of length $T$ is bounded by $NT/(\min_{i}r_iT)$.
\hfill $\blacksquare$

\begin{remark}\label{remarkchoicer}{\em 
Note that the conditions of Lemma \ref{nonzeno} rule out the case where there is a certain $r_i$ taking the value $0$. We do so because it could generate Zeno solutions. Specifically, if two agents $i$ and $j$ with bi-directional links have their tuning parameters $r_i$ and $r_j$ equal to $0$, there exists a solution in which agent $i$ resets $\tau_i$ from $1$ to $0$, triggering agent $j$ to update $\tau_j$ to $1$, which will be followed by an update of the form $\tau^+_j=0$, which in turn will trigger agent $i$ to update its state $\tau_i$ to $1$. The process repeats infinitely, generating a purely discrete-time solution. In order to avoid this behavior, we will introduce later in Theorem \ref{thm:specialcaseforzeror} a class of digraphs for which Zeno solutions do not emerge even when $r=0$.}  
\end{remark}

An advantage of formulating the closed-loop system of PCOs as HDS satisfying the basic conditions is that we can harness existing theoretical tools to establish suitable robustness results. Specifically, we have the following fact:  
\vspace{-0.2cm}
\begin{lemma}\label{robustness_lemma}
If $(r,\mathcal{G})$ is a sync-pair, then there exists a $\beta\in\mathcal{K}\mathcal{L}$ such that for each $\nu>0$ there exists $e^*>0$ such that for all measurable functions $e_i:\text{dom}(e)\to\mathbb{R}^N$, $i\in\{1,2,\ldots,6\}$, satisfying $\sup_{t\geq0}|e_i(t)|\leq e^*$, every solution of the perturbed HDS $\mathcal{H}(r, \mathcal{G})+e$, given by
\begin{subequations}\label{HDS_perturbed}
	\begin{align}
	&\tau+e_1\in C,~~~~~~~~~\dot{\tau}= f(\tau+e_2)+e_3,\label{HDS_flows_perturbed}\\
	&\tau+e_4\in D,~~~~~~\tau^+\in G(\tau+e_5)+e_6.~~\label{HDS_jumps_perturbed}
	\end{align}
\end{subequations}
satisfies the bound $|\tau(t,j)|_{\mathcal{A}_s}\leq \beta (|\tau(0,0)|_{\mathcal{A}_s},t+j)+\nu$, for all $(t,j)\in\text{dom}(\tau)$.

\end{lemma}
\textbf{Proof}: 
The result follows directly from item (b) of Def.~\ref{compatibility}, the compactness of $\mathcal{A}_s$, $C$, and $D$; the fact that $\mathcal{H}(r, \mathcal{G})$ is well posed, and the application of \cite[Lemma 7.20]{teel}. \hfill $\blacksquare$

\subsection{Scalability Issue and Negative Results}
We start by presenting a known result established in \cite[Thm. 1]{AHDS15} and \cite[Prop. 1]{sync_hybrid_poveda}: 
\begin{lemma}\label{lemma_1}
Let $r\in (0, \nicefrac{1}{N})^N$ and $\mathcal{G}$ be a strongly connected digraph. Then, $(r, \mathcal{G})$ is a sync-pair.  
Moreover, the set $\mathcal{A}_s$ is UGFxTs. Every maximal solution $\tau$ satisfies $|\tau(t,j)|_{\mathcal{A}_s}=0,~~~~\forall~~t\geq T^* := T$, with $(t,j)\in\text{dom}(\tau)$.
\end{lemma}

\begin{remark}{\em 
Although Lemma \ref{lemma_1} is a positive result, the condition $r\in (0, \nicefrac{1}{N})^N $ causes the scalability issue; indeed, since each $r_i$ is upper bounded by a term of order $\mathcal{O}(\frac{1}{N})$, the partition $[0,r_j)$ in \eqref{update_neigbors} associated to each agent vanishes as $N\to\infty$. Moreover, if the size of the network increases, in order to achieve fixed-time synchronization one would need to persistently re-tune the parameters $r_i$ of existing agents. While previous results in the literature \cite{Friedmanglobal} \cite{Nishimura} have presented scalable algorithms with partition parameters $r_i=\frac{1}{2}$ for all $i = 1,\ldots, N$, the synchronization results are only local or almost global.}
\end{remark}

By taking a closer look at the proof of Lemma~\ref{lemma_1}, we can relax slightly the condition by requiring that $r\in (0, \nicefrac{1}{N - 1})^N$. The trade-off is that the convergence time will be doubled as $2T$. However, such relaxation is still insufficient for fixing the scalability issue. 
Whether or not the condition can further be relaxed for some particular digraph is unknown.  Nevertheless, we present an impossibility result for the family of strongly connected digraphs.    

\vspace{-0.2cm}
\begin{proposition}\label{lem:strongconnected1}
The following holds for the HDS $\mathcal{H}(r,\mathcal{G}):=\{C,f,D,G\}$ given by \eqref{eq:DHDS}:
\begin{enumerate}[(a)]
    \item For every strongly connected digraph $\mathcal{G}$, and every $r\in(0,\frac{1}{N-1})^N$, the pair $(r,\mathcal{G})$ is a sync-pair.
    \item There exists a strongly connected digraph $\mathcal{G}$ such that for any $r\in (\frac{1}{(N - 1)}, 1]^N$, the pair $(r, \mathcal{G})$ is not a sync-pair. 
\end{enumerate}
\end{proposition}
\begin{figure}[t!]
	\centering
	\scalebox{0.09}{\subfloat{
    \centering
    \includegraphics[width=\textwidth]{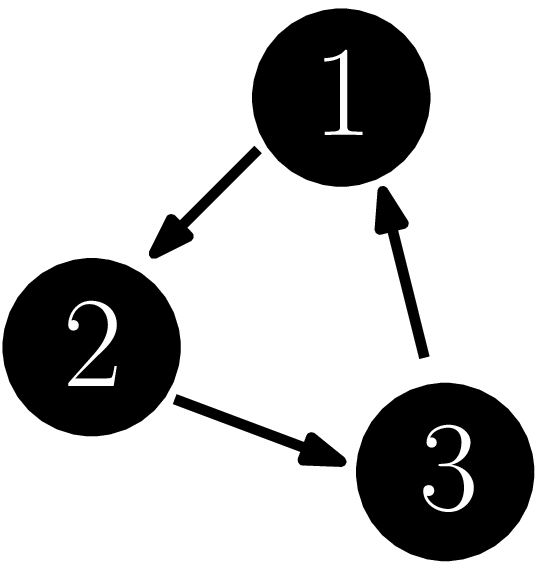}}}\hspace{1em}
	\scalebox{0.2}{\subfloat{
		\centering
		\includegraphics[width=\textwidth]{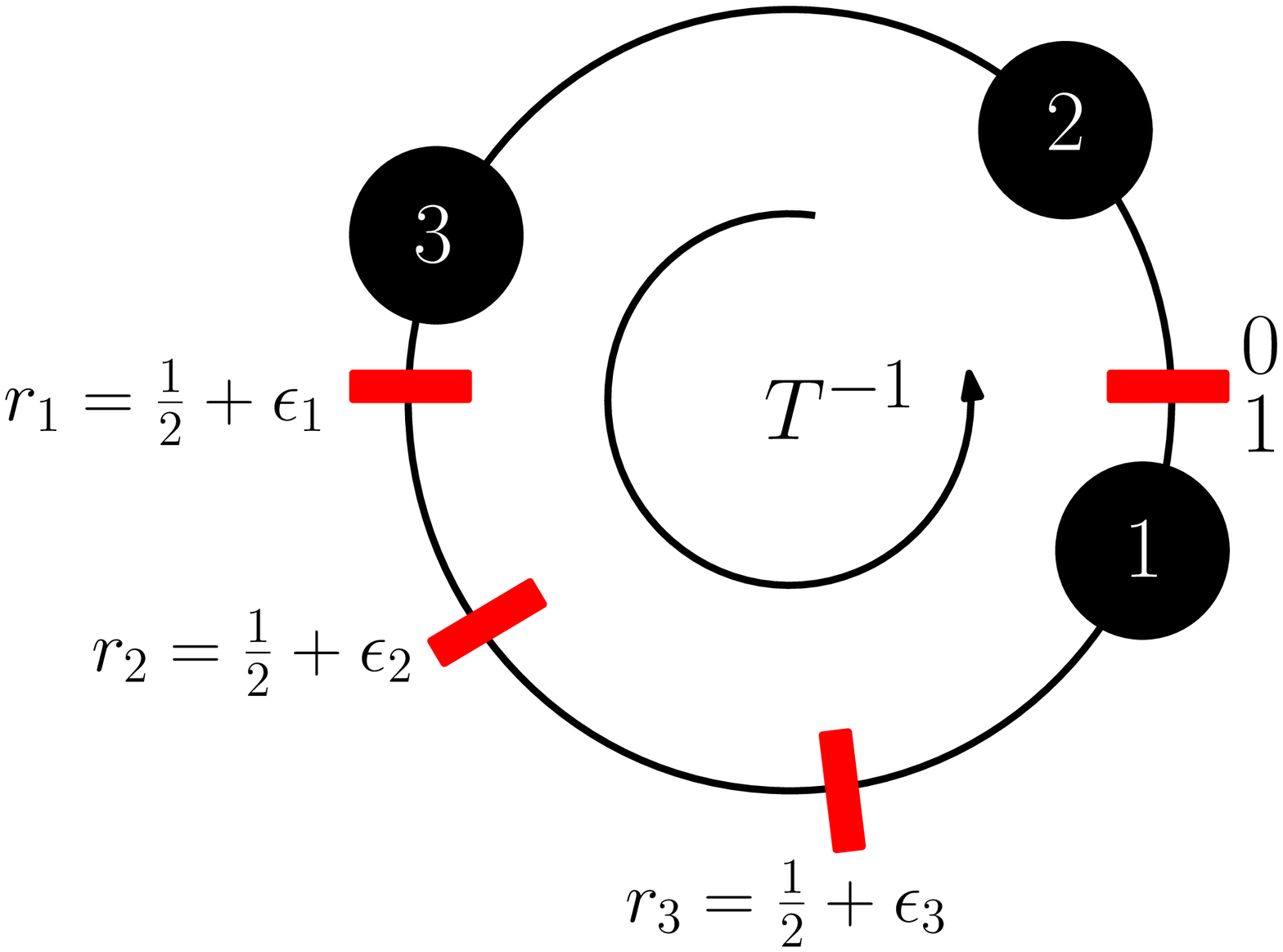}}}
    \caption{Left: A cycle digraph with $N=3$. Right: Representation of the PCOs in the unit circle with $r_i=\frac{1}{N-1}+\epsilon_i$, $\tau_1(0,0)=1-\epsilon$, $\tau_2(0,0)\in[0,r_2-\epsilon)$, and $\tau_3(0,0)=\frac{1}{2}-\epsilon$, considered in the proof of Proposition \ref{lem:strongconnected1}.}
    \label{fig:prop1}
\end{figure}
\vspace{-0.2cm}
\textbf{Proof}: We first show the existence of a uniformly bounded time $t^*\geq0$ such that every solution of $\mathcal{H}(r,\mathcal{G})$ satisfies $\tau(t^*,0)\in D$ for any $r\in[0,1)$ and any digraph $\mathcal{G}$. Indeed, let $r\in[0,1)$ and $\mathcal{G}$ be given, and consider a solution of $\mathcal{H}(r,\mathcal{G})$. Let $\tau(0,0)\in C\backslash D$, otherwise there is nothing to prove. Then, it must be the case that $\tau_i(0,0)\in[0,1)$ for all $i\in\mathcal{V}$. By the construction of the flow map $f$ in \eqref{flow_map_deterministic}, it follows that during flows the solutions satisfy $\tau_i(t,j)=\frac{1}{T}(t-t_j)+\tau_i(t_j,j)$ for all $i\in\mathcal{V}$, where $t_j:=\inf\{t\geq0:(t,j)\in\text{dom}(\tau)\}$. Since the function is increasing in $t$, setting $j=0$ and $t_j=0$, there must exist $i^*\in\mathcal{V}$ and $t^*\geq 0$ such that $\tau_{i^*}(t^*,0)=1$. In turn, this implies

\vspace{-0.6cm}
\begin{equation}\label{bound_first_jump}
t^*=T(1-\tau_{i^*}(0,0))\leq T,
\end{equation}

\vspace{-0.4cm}
for all $\tau_i\in[0,1)$. Now, to establish item (a), note that by Lemma \ref{nonzeno} every solution of the system is complete and uniformly non-Zeno. To show UGAS of $\mathcal{A}_{s}$, let us consider the Lyapunov function $V:[0,1]^N\to\mathbb{R}_{\geq0}$ defined as the infimum of all arcs that touch all agents on the unit circle, where the points $0$ and $1$ are identified to be the same point. Note that this function is positive definite with respect to $\mathcal{A}_s$, and by construction $V$ satisfies $0\leq V(\tau(0,0))\leq 1-\frac{1}{N}$ for every possible initial condition $\tau(0,0)\in[0,1]^N$. Moreover, since during flows the function $V$ does not change, and during jumps the function $V$ cannot increase, it follows that the bound $V(\tau(t,j))\leq 1-\frac{1}{N}$ holds for all solutions $\tau$ of the HDS $\mathcal{H}(r,\mathcal{G})$. Additionally, by equation \eqref{bound_first_jump} we know that every solution will experience a jump at the hybrid time $(t^*,0)$, triggered by at least one agent $i^*\in\mathcal{V}$ satisfying $\tau_{i^*}=1$. Since the digraph is strongly connected, and the update rule \eqref{update_neigbors} is binary, there exists at least one agent $j^*$ that is an out-neighbor of agent $i$ such that $\tau_{j^*}(t^*,1)\in\{0,1\}$. Therefore, agents $i^*$ and $j^*$ are now at the same position on the unit circle, and it follows that $0\leq V(\tau(t^*,1))\leq 1-\frac{1}{N-1}$. From this point, by using the same arguments that lead to equation \eqref{bound_first_jump}, the system can stay in the flow set for at most $T^*$ seconds, which implies the existence of an agent $i^{**}$ and a time $t^{**}>0$ such that $\tau_{i^{**}}(t^{**},1)=1$. Since $V(\tau(t^{**},1))\leq 1-\frac{1}{N-1}$ still holds, it follows that necessarily  $\tau_i\geq\frac{1}{N-1}$ for all $i\in\mathcal{V}$, and since the tuning parameters of all agents satisfy $r_i\in(0,\frac{1}{N-1})$, it follows that $\tau_i\in(r_i,1]$ for all $i\in\mathcal{V}$, and by \eqref{update_neigbors} the system will experience $N-1$ consecutive jumps, after which $\tau_i(t^{**},j^*)=0$ for all $i\in\mathcal{V}$. This establishes finite-time synchronization of the network with $t^{**}\leq 2T$, which, in turn, implies that there is no complete solution that keeps the Lyapunov function $V$ in a non-zero level set. By the hybrid invariance principle \cite[Thm. 8.8]{teel} we conclude UGAS of $\mathcal{A}_s$. Finally, note that since every solution is uniformly non-Zeno, and the set of initial conditions is compact, there exists $j^*>0$ such that the HDS will experience at most $j^*$ jumps in any continuous-time interval of length $2T$. Therefore, the set $\mathcal{A}_s$ is actually UGFxTS with $\bar{T}=2T+j^*$.

\begin{figure}[t!]
    \centering
    \includegraphics[width=0.2\textwidth]{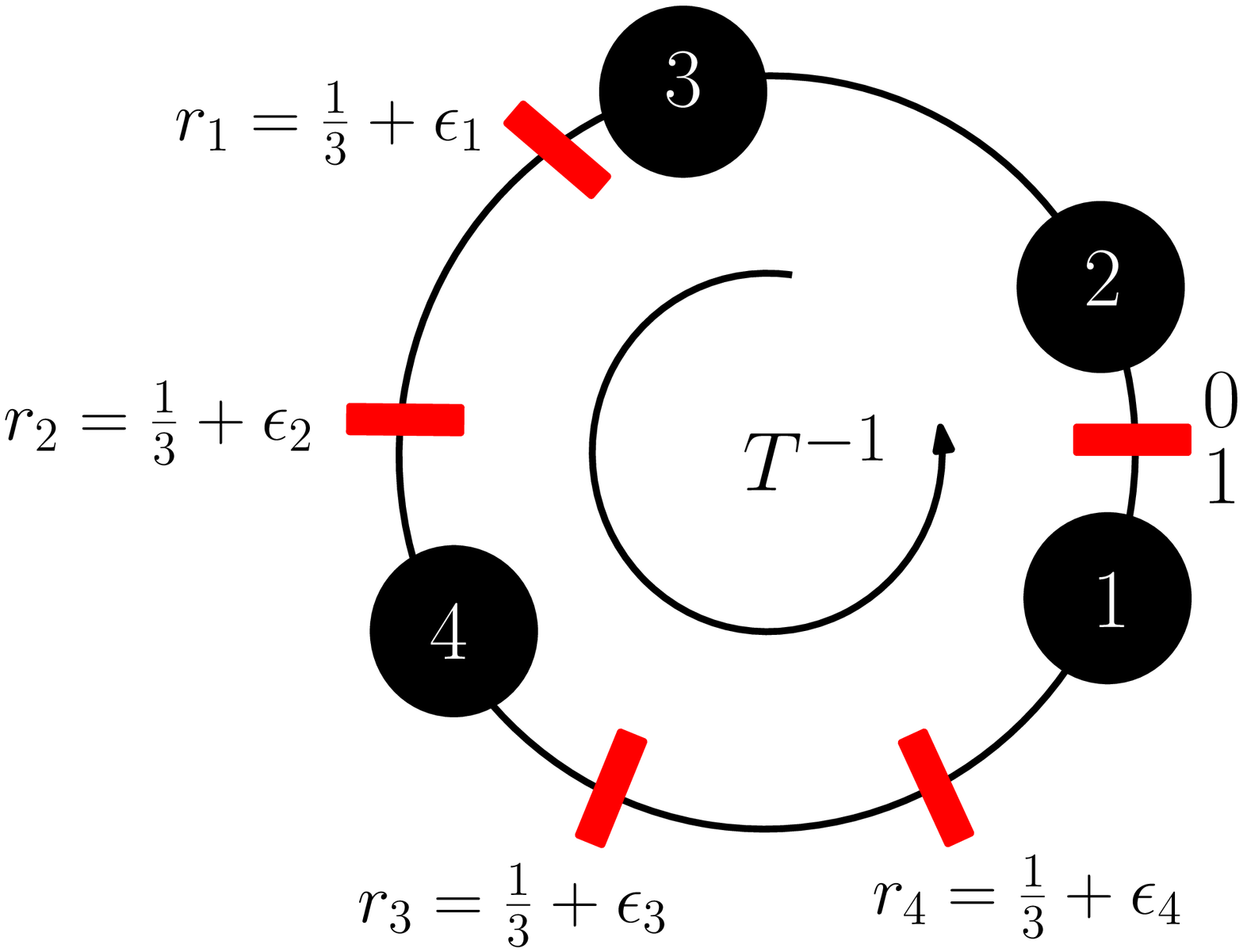}~~~~~~\includegraphics[width=0.22\textwidth]{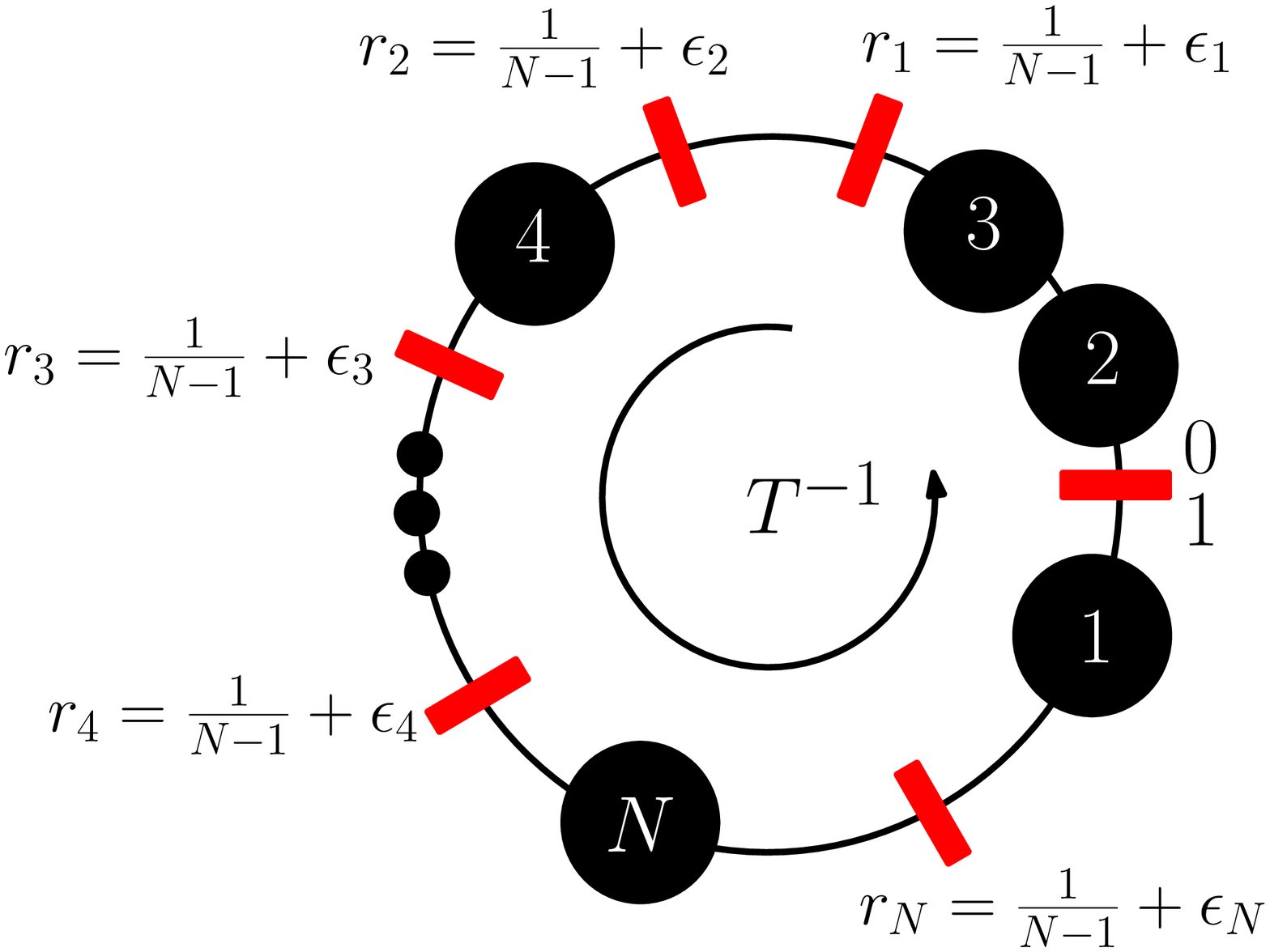}
    \caption{Problematic sets of initial conditions for PCOs on cycle digraphs, with local partitions $r_i=\frac{1}{N-1}+\epsilon_i$. Left: N=3. Right: Arbitrary $N>0$.}
    \label{fig:prop2}
\end{figure}

To prove item (b), it suffices to consider a counter-example. Consider the cycle digraph shown in Figure \ref{fig:prop1}-(a) having three vertices. Let $r\in(\frac{1}{2},1]^N$ be given. Define $\tau_1(0,0):=\tau_{1,0}$, $\tau_2(0,0):=\tau_{2,0}$ and $\tau_3(0,0):=\tau_{3,0}$. Without loss of generality we take $T=1$. Choose the initial conditions such that the following inequalities hold: (a) $\tau_{10}>\frac{1}{2}$; (b) $\tau_{3,0}=\tau_{1,0}-\frac{1}{2}$; (c) $0<\tau_{2,0}<\tau_{1,0}-(1-r_2)$. By the properties of $r_i$ it follows that $\tau\in(0,1]^N$ and $\tau_{1,0}>\tau_{i,0}$, $i\in\{2,3\}$. Moreover, 

\vspace{-1cm}
\begin{equation}\label{useful_inequaitiesprop1}
\tau_{1,0}-\tau_{3,0}<r_1,~~\tau_{1,0}-\tau_{3,0}>1-r_3,~~\tau_{1,0}-\tau_{3,0}<r_2.
\end{equation}

\vspace{-0.5cm}
and
\begin{equation}\label{equalityhalf}
\tau_{3,0}+(1-\tau_{1,0})=\tau_{1,0}-\tau_{3,0}.
\end{equation}
We now proceed to track the behavior of the solutions generated by the HDS from the given initial conditions. In particular, after $1-\tau_{1,0}$ seconds of flow, the system will satisfy the condition 
\vspace{-1cm}

\begin{equation}\label{periodic00}
\tau_1=1,~~\tau_2=\tau_{2,0}+(1-\tau_{1,0}),~~\tau_3=\tau_{3,0}+(1-\tau_{1,0}),
\end{equation}

\vspace{-0.5cm}
which, by property (c), triggers the jumps $\tau_1^+=0$, $\tau_2^+=0$, and $\tau_3^+=\tau_3$. From here, the system will flow for $\tau_{1,0}-\tau_{3,0}$ seconds, until the following condition holds:

\vspace{-0.7cm}
\begin{equation*}
\tau_1=\tau_{1,0}-\tau_{3,0},~~\tau_2=\tau_{1,0}-\tau_{3,0},~~\tau_3=1,
\end{equation*}
which, by the first inequality in \eqref{useful_inequaitiesprop1}, triggers the jumps $\tau_1^+=0$, $\tau_2^+=\tau_2$, and $\tau_3^+=0$. The system will now flow for $\tau_{3,0}+(1-\tau_{1,0})$ seconds, until the following conditions hold:

\vspace{-1cm}
\begin{equation*}
\tau_1=\tau_{3,0}+(1-\tau_{1,0}),~~\tau_2=1,~~\tau_3=\tau_{3,0}+(1-\tau_{1,0}),
\end{equation*}

\vspace{-0.6cm}
which, by the second inequality in \eqref{useful_inequaitiesprop1}, triggers the jumps $\tau_1^+=\tau_1$, $\tau_2^+=0$, and $\tau_3^+=0$. From here the system will flow for $\tau_{1,0}-\tau_{3,0}$ seconds, until the following conditions hold:

\vspace{-0.8cm}
\begin{equation*}
\tau_1=1,~~\tau_2=\tau_{1,0}-\tau_{3,0},~~\tau_3=\tau_{1,0}-\tau_{3,0},
\end{equation*}
which, by the third condition in \eqref{useful_inequaitiesprop1}, trigger jumps $\tau_1^+=0$, $\tau_2^+=0$ and $\tau_3^+=\tau_3$. Using \eqref{equalityhalf} we can observer that this is the exact same state obtained after \eqref{periodic00}. Therefore, the system has entered a periodic solution that does not contain points of the set $\mathcal{A}_s$, i.e., synchronization is never achieved. The counter-example can be generalized to any dimension $N$ as shown in Figure \ref{fig:prop2}.  
\hfill $\blacksquare$ 

Although the proof of the negative result of Proposition \ref{lem:strongconnected1} is built upon cycle digraphs, there are other strongly connected digraphs that also have the  scalability issue. However, a complete characterization of these digraphs still remains open.  
To fix the scalability issue, we consider below digraphs beyond the strongly connected ones. We start with the following necessary condition:     

\vspace{-0.2cm}
\begin{proposition}\label{lem:rooted_necessity}
If $(r, \mathcal{G})$ is a sync-pair, then $\mathcal{G}$ is rooted.  
\end{proposition}

\vspace{-0.2cm}
\textbf{Proof}: 
We apply strong component decomposition (see, for example,~\cite{chen2017controllability}) to $\mathcal{G}$ and obtain strongly connected subgraphs $\mathcal{G}_l = (\mathcal{V}_l, \mathcal{E}_l)$, for $l = 1,\ldots, k$, where the subsets $\mathcal{V}_l$ form a partition of $\mathcal{V}$. 
A subgraph $\mathcal{V}_l$ is said to be a leading strong component if for any $v_j\in \mathcal{V}\backslash\mathcal{V}_l$ and any $v_i\in \mathcal{V}_l$, $(i,j)$ is not an edge of $\mathcal{G}$. If a digraph $\mathcal{G}$ is not rooted, then it has at least two leading components. Without loss of generality, we assume that $\mathcal{G}_1$ and $\mathcal{G}_2$ are two leading components. 
Because a leading component does not have incoming neighbors, its dynamics are completely decoupled from the others. In particular, the dynamics of components $\mathcal{G}_1$ and $\mathcal{G}_2$ are independent of each other. Thus, the overall system cannot achieve synchronization from all initial conditions.   \hfill $\blacksquare$ 

\begin{figure}[t!]
    \centering
    \includegraphics[width=0.11\textwidth]{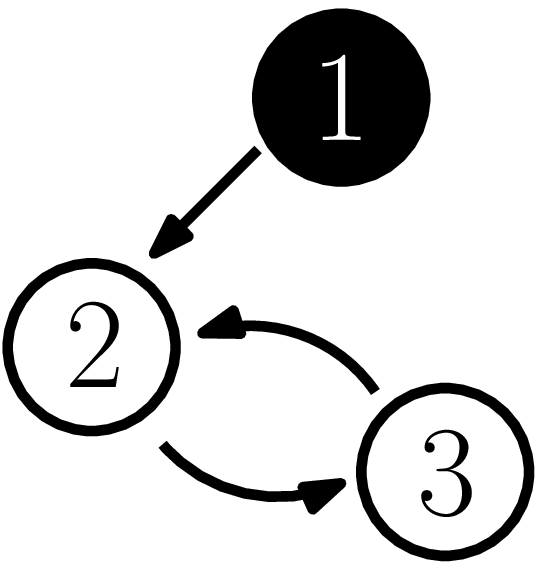}~~~\includegraphics[width=0.15\textwidth]{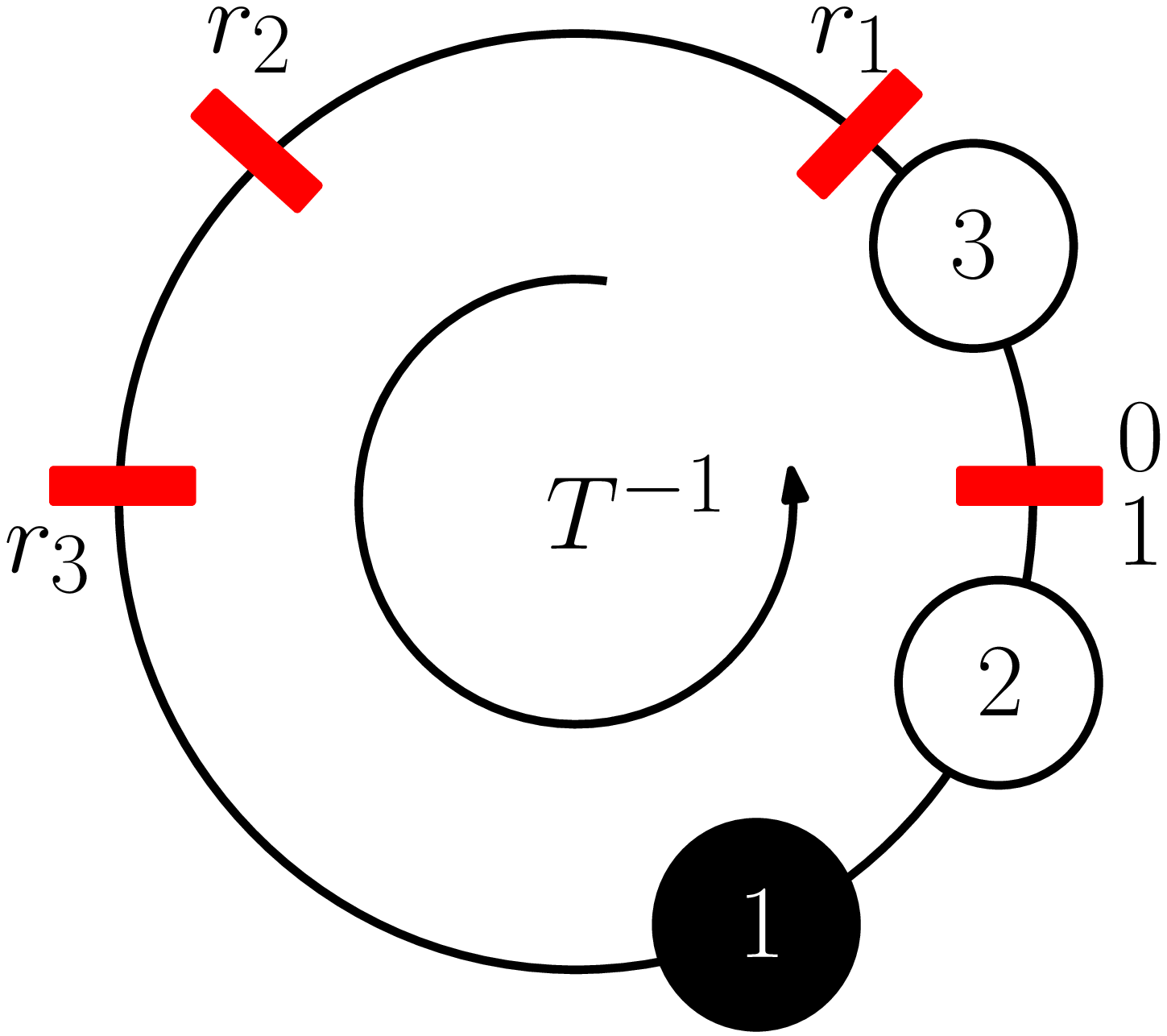}~~~~\includegraphics[width=0.15\textwidth]{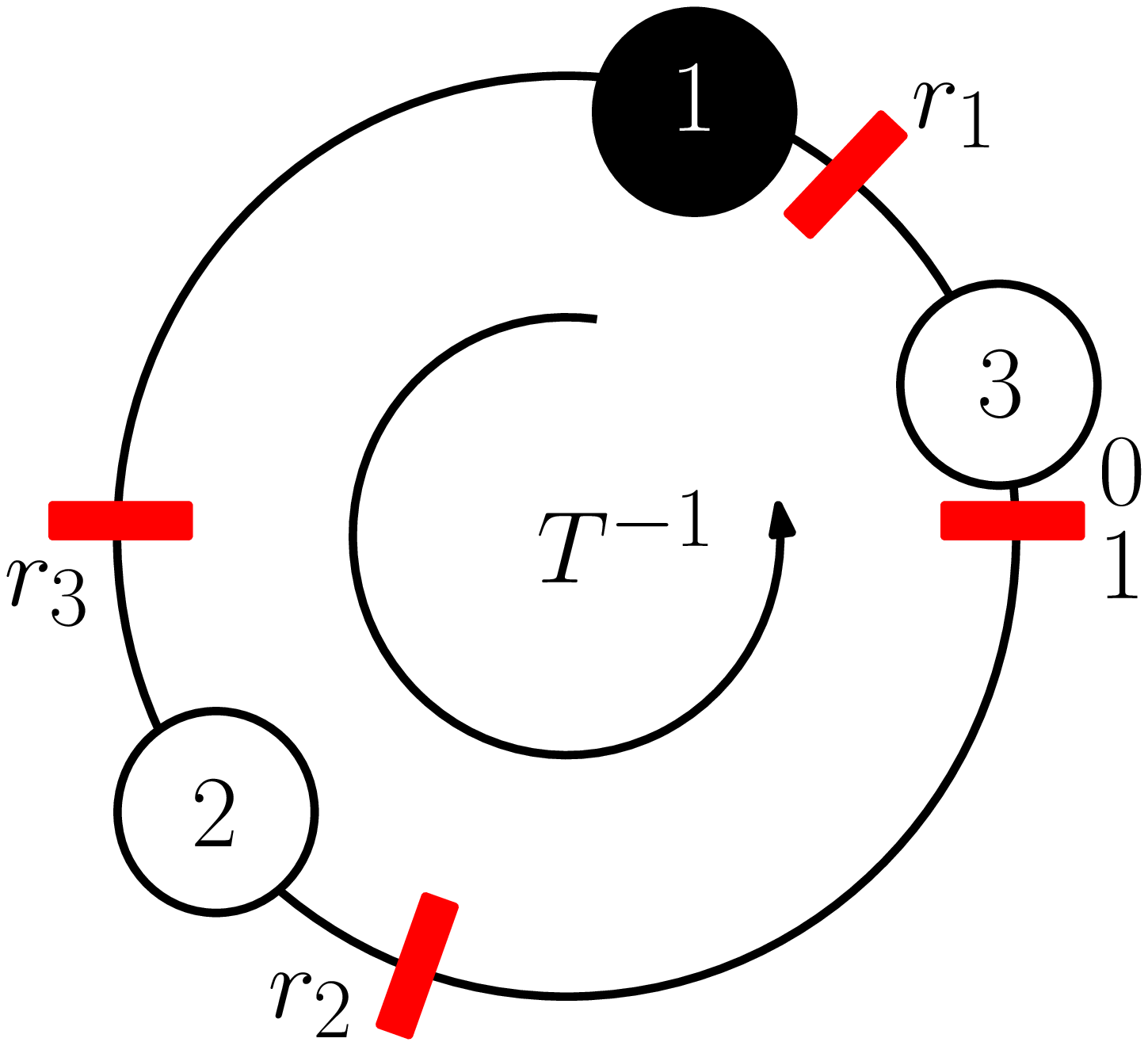}\caption{Graph and initial conditions for the counter-example considered in the proof of Proposition \ref{prop:rooted_notsufficient}. Left: Rooted digraph. Center: Problematic initialization for $r_2\leq 0.5$. Right: Problematic initialization for $r_2\geq 0.5$ }
    \label{fig:prop3}
\end{figure}

%
\vspace{-0.2cm}
Conversely, we can ask whether for any given rooted digraph $\mathcal{G}$, there is a partition vector $r$ such that $(r, \mathcal{G})$ is a sync-pair? The following result provides a negative answer: 

\vspace{-0.2cm}
\begin{proposition}\label{prop:rooted_notsufficient}
 There exists a rooted digraph $\mathcal{G}$ such that for any $r\in (0, 1]^N$, the pair $(r, \mathcal{G})$ is not a sync-pair.
\end{proposition}
\vspace{-0.2cm}
\textbf{Proof:} The proof is constructive. 
Let us consider the digraph shown in Figure \ref{fig:prop3}, and suppose that $r\in(0,1]^N$. Without loss of generality we consider $T=1$. Define $\tau_{i,0}=\tau_i(0,0)$ for $i\in\{1,2,3\}$. We consider two scenarios: (a) $r_2\leq 0.5$; and (b) $r_2>0.5$. 
\emph{Scenario (a):} Choose $\tau_{2,0}$ such that $1-r_3<\tau_{2,0}\leq 1$. Choose $\tau_{1,0}$ such that

\vspace{-0.5cm}
\begin{equation}\label{inequality2}
\max\{0,\tau_{2,0}-r_2\}<\tau_{1,0}< \tau_{2,0},
\end{equation}

\vspace{-0.2cm}
and choose $\tau_{3,0}$ such that

\vspace{-0.5cm}
\begin{equation}\label{inequality3}
0<\tau_{3,0}< \min \{\tau_{1,0},\tau_{2,0}-(1-r_3)\}.
\end{equation}
\noindent Note that this initialization satisfies $1>\tau_{2,0}>\tau_{1,0}>\tau_{3,0}>0$. Following the hybrid dynamics we obtain the following sequence of events:  Agents flow for $1-\tau_{2,0}$ seconds, until the system satisfies the condition $\tau_1=\tau_{1,0}+(1-\tau_{2,0})$,$\tau_2=1$, $\tau_3=\tau_{3,0}+(1-\tau_{2,0})$. By the right hand side of inequality \eqref{inequality3} we obtain that $\tau_3<r_3$, and the states will jump as $\tau_1^+=\tau_1$, $\tau_2^+=0$ and $\tau_3^+=0$. Following this jump, the system will flow for $1-(\tau_{1,0}+(1-\tau_{2,0}))$ seconds, until the system satisfies the following condition:

\vspace{-0.6cm}
\begin{equation}\label{periodic_state}
\tau_1=1,~~\tau_2=\tau_{2,0}-\tau_{1,0},~\tau_3=\tau_{2,0}-\tau_{1,0}.
\end{equation}

\vspace{-0.2cm}
\noindent By the left hand side of inequality \eqref{inequality2} we have $\tau_2<r_2$. Therefore, the system will jump as $\tau_1^+=0$, $\tau_2^+=0$ and $\tau_3^+=\tau_3$. Following this jump, the system will flow for $1-\tau_{2,0}+\tau_{1,0}$ seconds, until the following condition holds:

\vspace{-1cm}
\begin{equation}\label{boundbprop3}
\tau_1=\tau_{1,0}+(1-\tau_{2,0}),~\tau_2=\tau_{1,0}+(1-\tau_{2,0}),~\tau_3=1.
\end{equation}

\vspace{-0.6cm}
\noindent Since $r_2\leq 0.5$, it follows that $r_2\leq 1-r_2$ and $\tau_{0,2}-r_2\geq \tau_{2,0}-(1-r_2)$. Therefore, using again the left hand side of inequality \eqref{inequality2} we obtain that $\tau_{2}>r_2$ in \eqref{boundbprop3}. Thus, the system will jump as $\tau_1^+=\tau_1$, $\tau_2^+=1$ and $\tau_3^+=0$. At this point the system will jump again as $\tau_1^+=\tau_1$, $\tau_2^+=0$ and $\tau_3^+=0$. After $1-\tau_1$ seconds of flow, the system will satisfy the condition

\vspace{-0.6cm}
\begin{equation}\label{boundbprop3}
\tau_1=1,~\tau_2=\tau_{2,0}-\tau_{1,0},~\tau_3=\tau_{2,0}-\tau_{1,0},
\end{equation}

\vspace{-0.3cm}
\noindent which is the same state described in \eqref{periodic_state}, i.e., the system has entered a periodic cycle which does not include points in the set $\mathcal{A}$.

\emph{Scenario (b):}  Choose $\tau_{2,0}$ such that $\max\{1-r_3,1-r_2\}<\tau_{2,0}\leq 1$ holds. Choose $\tau_{1,0}$ such that

\vspace{-0.5cm}
\begin{equation}\label{inequality2b}
\max\{0,\tau_{2,0}-r_2\}<\tau_{1,0}< \tau_{2,0}-(1-r_2).
\end{equation}

\vspace{-0.2cm}
\noindent This choice is always possible given that $r_2>0.5$. Choose $\tau_{3,0}$ such that

\vspace{-0.6cm}
\begin{equation}\label{inequality3b}
0<\tau_{3,0}< \min \{\tau_{1,0},\tau_{2,0}-(1-r_3)\}.
\end{equation}

\vspace{-0.3cm}
\noindent Note that this initialization is always feasible and satisfies $1>\tau_{2,0}>\tau_{1,0}>\tau_{3,0}>0$. Following the hybrid dynamics we obtain the following sequence of events: 
Agents flow for $1-\tau_{2,0}$ seconds until the states satisfy $\tau_1=\tau_{1,0}+(1-\tau_{2,0})$,~$\tau_2=1$, $\tau_3=\tau_{3,0}+(1-\tau_{2,0})$. By the right hand side of inequality \eqref{inequality3b} we obtain that $\tau_3<r_3$, and the states will jump as $\tau_1^+=\tau_1$, $\tau_2^+=0$ and $\tau_3^+=0$. Following this jump, the system will flow for $\tau_{2,0}-\tau_{1,0}$ seconds, until the system satisfies the following condition:

\vspace{-0.6cm}
\begin{equation}\label{periodic_state2}
\tau_1=1,~~\tau_2=\tau_{2,0}-\tau_{1,0},~\tau_3=\tau_{2,0}-\tau_{1,0}.
\end{equation}

\vspace{-0.3cm}
\noindent By the left hand side of inequality \eqref{inequality2b} we have $\tau_2<r_2$. Therefore, the system will jump as $\tau_1^+=0$, $\tau_2^+=0$ and $\tau_3^+=\tau_3$. Following this jump, the system will flow for $1-\tau_{2,0}+\tau_{1,0}$ seconds, until the following condition holds:

\vspace{-0.9cm}
\begin{equation}\label{boundbprop3b}
\tau_1=\tau_{1,0}+(1-\tau_{2,0}),~\tau_2=\tau_{1,0}+(1-\tau_{2,0}),~\tau_3=1.
\end{equation}
By right-hand side of \eqref{inequality2b} we obtain that $\tau_{2}<r_2$ in \eqref{boundbprop3b}. Thus, the system will jump as $\tau_1^+=\tau_1$, $\tau_2^+=0$ and $\tau_3^+=0$. After $1-\tau_1$ seconds of flow, the system will satisfy the condition

\vspace{-0.7cm}
\begin{equation}\label{boundbprop3}
\tau_1=1,~\tau_2=\tau_{2,0}-\tau_{1,0},~\tau_3=\tau_{2,0}-\tau_{1,0},
\end{equation}

\vspace{-0.2cm}
which is the same state \eqref{periodic_state2}, i.e., the system has entered a periodic cycle which exclude points in the set $\mathcal{A}_s$.

Since scenarios (a) and (b) cover every possible choice of $r\in(0,1]^N$, and for any choice we found a solution that does not converge to $\mathcal{A}_s$, the pair $(r,\mathcal{G})$ is not a sync-pair for any partition vector $r\in(0,1]^N$. \hfill $\blacksquare$ 
\vspace{-0.2cm}
\begin{remark}{\em 
    It is well known that for standard Laplacian (consensus) dynamics in $\mathbb{R}^N$, the digraph $\mathcal{G}$ being rooted is a necessary and sufficient condition for state synchronization. Moreover, finite-time consensus in $\mathbb{R}^N$ can also be achieved under the same graphical condition \cite{WangFiniteSync}.
    Proposition~\ref{lem:rooted_necessity} shows that, in order to achieve synchronization, the connectivity requirement for the network of PCOs is completely different from the one for the standard Laplacian dynamics. However, we also note that the negative result, Proposition~\ref{prop:rooted_notsufficient}, holds only for the deterministic resetting algorithm with arbitrary initialization. Indeed, as shown in \cite[Thm. 2]{Anton17}, synchronization over rooted digraphs can be achieved if all the PCOs are initialized within a semi-circle. Later in the next section, we will show that $\mathcal{G}$ being rooted is actually sufficient for global synchronization using a stochastic resetting algorithm.      
    }
\end{remark}

\vspace{-0.2cm}
\subsection{Positive Result on Rooted Acyclic Digraphs}
\vspace{-0.2cm}
In this subsection, we focus on a special class of rooted digraphs, namely rooted acyclic digraphs. We will show that for every such digraph $\mathcal{G}$ and for every partition vector $r\in[0,1]^N$, the tuple  
$(r, \mathcal{G})$ is a sync-pair. In particular, the choice of $r$ can be made independent of the size $N$ of the digraph. We formulate the result in the following theorem: 
\vspace{-0.2cm}
\begin{theorem}\label{main_theorem1}
For any rooted acyclic digraph $\mathcal{G}$ and any  $r\in (0,1]^N$, $(r, \mathcal{G})$ is a sync-pair. Moreover, every maximal solution $\tau$ satisfies

\vspace{-0.9cm}
\begin{equation}\label{finite_time}
|\tau(t,j)|_{\mathcal{A}_s}=0,~~~~\forall~~t\geq T^*:=(\dep(\mathcal{G})+1)T,
\end{equation}

\vspace{-0.5cm}
with $(t,j)\in\text{dom}(\tau)$. 
\end{theorem}
\vspace{-0.2cm}
\textbf{Proof:} We consider again the Lyapunov function $V:[0,1]^N\to \mathbb{R}_{\geq0}$ defined as the infimum of all the arcs that touch all agents on the unit circle, where the points $0$ an $1$ are identified to be the same. By \cite{AHDS15}, this Lyapunov function satisfies the following properties: (i) It is positive definite with respect to the compact set (2). (ii) It remains constant during flows because all the oscillators have the same frequency $\frac{1}{T}$. (iii) It does not increase at jumps since  jumps never increase the number of distinct points occupied by the agents.  We claim that there is no maximal solution of the HDS $\mathcal{H}(r,\mathcal{G})$ that keeps $V$ equal to a non-zero constant. We show this by establishing fixed-time synchronization. Let $\tau(0,0)\in[0,1]^N$ and $\tau$ be a solution of the HDS \eqref{eq:DHDS}. 
Recall that $\mathcal{V}_l$ defines all vertices/agents of depth~$l$. Since the digraph is rooted acyclic, no agent can influence the unique root agent, and without loss of generality, we assume that $\tau_1$ corresponds to the root agent, i.e., $\mathcal{V}_0 =\{\tau_1\}$. Based on this, we proceed to establish a uniform lower bound on the amount of hybrid time that can pass before the Lyapunov function is exactly equal to zero. We establish the fact by induction on the depth $l$ of the vertices of $\mathcal{G}$. 
\vspace{-0.2cm}
\begin{itemize}  
\item \textit{Base Case $l = 1$:}  In at most $T$ seconds of flow, $\tau$ will satisfy $\tau_{1}=1$, and agent $1$ will trigger all the vertices on $\mathcal{V}_1$ to either jump to 0 or 1. Thus, based on $r$, there will exist a partition of $\mathcal{V}_1$ that is defined by the index sets  ($I', I'',I''' $) such that: (i) for all $i' \in I'$, $\tau_{i'} > r_i$ (the agents $i'$ will jump to 1 and trigger $\mathcal{V}_2$); (ii) for all $i'' \in I''$, $\tau_{i''} < r_i$ (the agents $i''$ will jump to 0 and flow for at most $T$ seconds to trigger $\mathcal{V}_2$); (iii) for all $i''' \in I'''$, $\tau_{i'''} = r_i$ (the agents $i'''$ will have a set-valued jump $\{0,1\}$. If the agent jumps to 1, it will follow (i), otherwise, it will follow (ii)). Note that after the first jump, $\mathcal{V}_0$ synchronize with $\mathcal{V}_1$ within at most $2T$ seconds and remain synchronized since $\mathcal{V}_1$ does not influence $\mathcal{V}_0$ by the acyclic property of the digraph. 
\item \textit{Induction Step:} Suppose that agents $\{\mathcal{V}_0,\mathcal{V}_1,\cdots, \mathcal{V}_k\}$ synchronize in at most  $(k+1)T$ seconds, where $k<\dep(\mathcal{G})$. Since the digraph does not have a cycle, and the root/agent has a path to all the agents, we have that agents $\mathcal{V}_k$ only influence agents $\mathcal{V}_{k+1}$ and cannot affect already synchronized agents $\mathcal{V}_{l}$, for $0\le l \le k - 1$. 
Thus, agents $\{\mathcal{V}_0,\mathcal{V}_1,\cdots, \mathcal{V}_{k+1}\}$ synchronize in at most $(k+2)T$ seconds. Therefore, the agents $\{\mathcal{V}_0,\mathcal{V}_1,\cdots, \mathcal{V}_{\dep(\mathcal{G})}\}$ synchronize in at most  $(\dep(\mathcal{G})+1)T$ seconds and remain synchronized after that, i.e., they occupy the same position on the unit circle for all $(t,j)\in\text{dom}(\tau)$.
\end{itemize}
\vspace{-0.2cm}
Furthermore, by Lemma \ref{nonzeno}, the above arguments imply that $V(\tau(t,j))=0$ for all $(t,j)\in\text{dom}(\tau)$ such that $t+j\geq(\dep(\mathcal{G})+1)(T+N\underline{r}^{-1})=:\overline{T}$. Since $\tau$ was arbitrary, we have established that there is no solution of the HDS that keeps the Lyapunov function in a non-zero level set. We can now directly establish UGAS of the HDS $\mathcal{H}(r,\mathcal{G})$ with respect to the compact set $\mathcal{A}_s$ by using the Hybrid Invariance Principle \cite{invariance}. Absence of purely or eventually discrete-time solutions follows by Lemma \ref{nonzeno}. This completes the proof of the Theorem. \hfill $\blacksquare$
\begin{figure}[t!]
    \centering
    \includegraphics[width=0.12\textwidth]{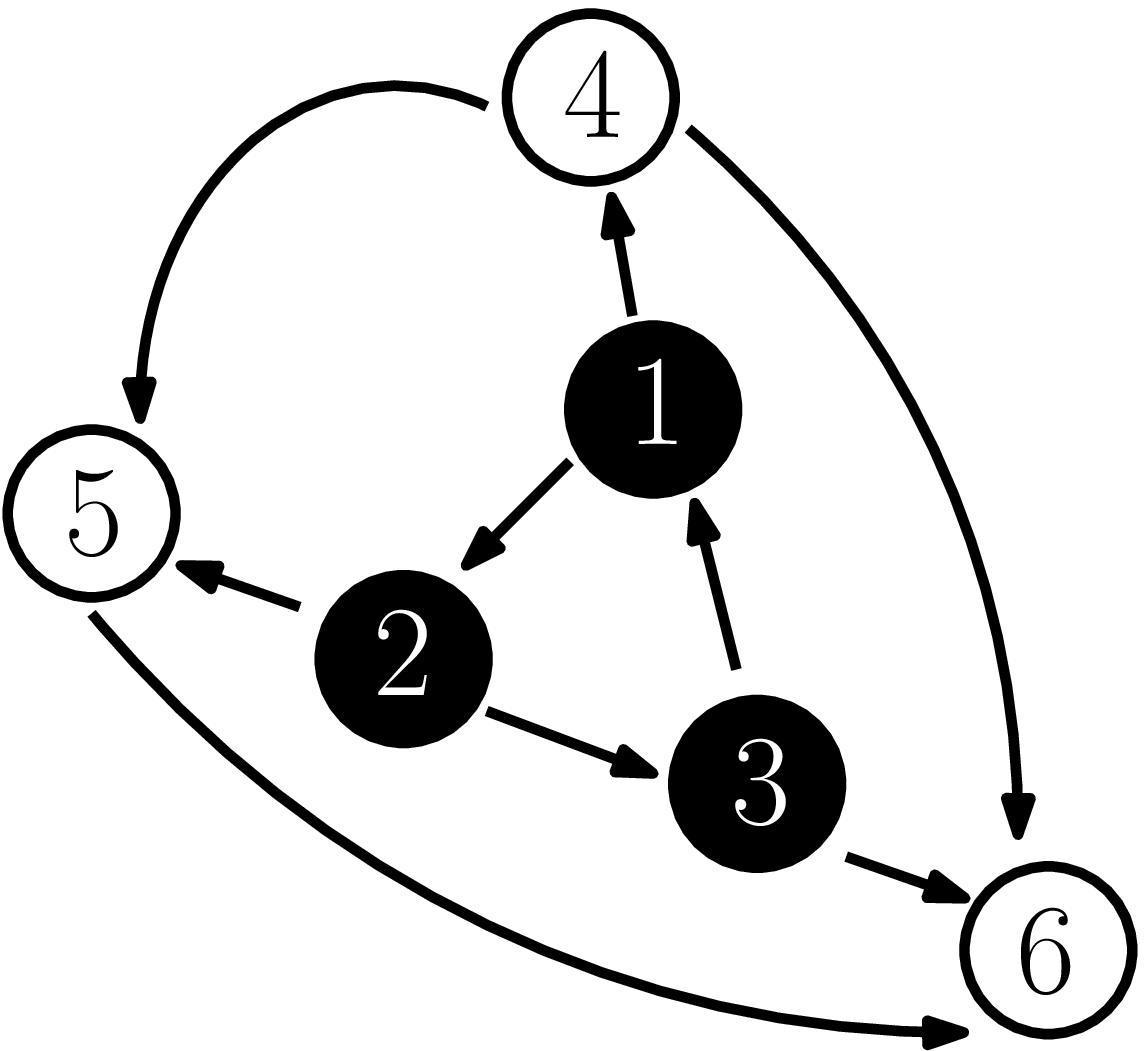}~~~~~~~~~~~~\includegraphics[width=0.11\textwidth]{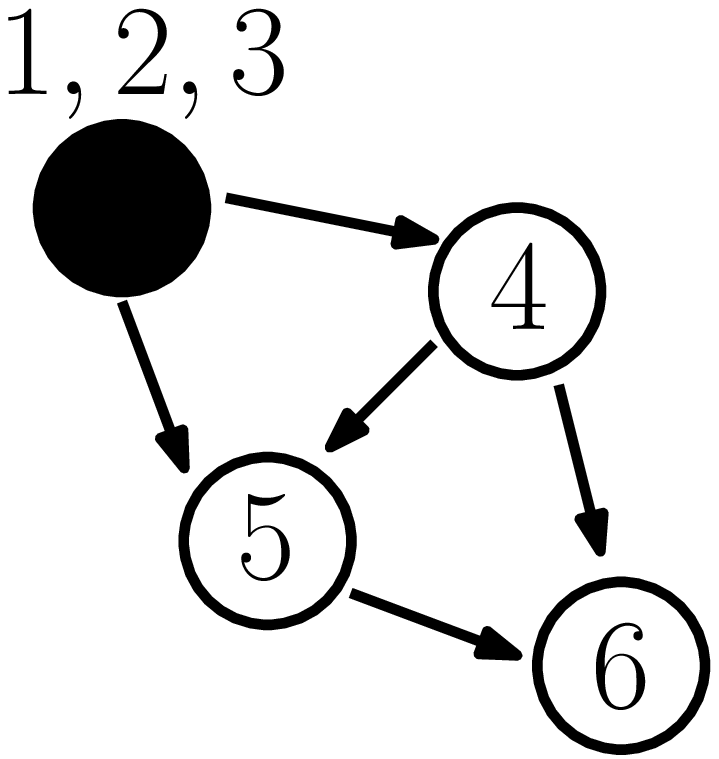}
    \caption{(a) Quasi-acyclic digraph $\mathcal{G}$. (b) Condensed digraph $\mathcal{G}_c$.}
    \label{fig:rootcondensation}
\end{figure}
\vspace{-0.1cm}

For the special case where $r = 0_N$, we have the following:
\vspace{-0.2cm}
\begin{theorem}\label{thm:specialcaseforzeror}
A pair $(0_N, \mathcal{G})$ is a sync-pair if and only if $\mathcal{G}$ is rooted acyclic. In this case,  \eqref{finite_time} holds with $T^*:=T$.
\end{theorem}
\vspace{-0.2cm}
\textbf{Proof:}
\emph{Sufficiency:} First, we show that every solution is non-Zeno. Indeed, by construction, Zeno behavior can only occur if there exists a solution $\tau$ that remains in the jump set $D$ for all $(t,j)\in\text{dom}(\tau)$. In order to remain in $D$, for such solution there must exist an agent $i$ satisfying $\tau_i(0,0)=1$, and an agent $j$ and a path from $i$ to $j$ and from $j$ to $i$. Otherwise, after at most $N$ jumps all other agents have already been triggered to $0$ and $\tau\notin D$. However, since by assumption the digraph is acyclic, there are no two vertices $i,j$ that have a path from each other. Thus, at most $N$ consecutive jumps can occur in the system until the state satisfies $\tau=0_N\in C\backslash D$. To show UGAS, note that since the root agent is not affected by any other vertex, for every solution of the HDS $\mathcal{H}(0_N,\mathcal{G})$ there exists $(t^*,j^*)\in\text{dom}(\tau)$ with $t^*\leq T$ and $j^*\leq N$ such that the state of the root vertex $v^*$ satisfies $\tau_{v^*}(t^*,j^*)=1$. Based on this, we claim that every solution $\tau$ will satisfy $V(\tau(t,j))=0$ for all $(t,j)\in\text{dom}(\tau)$ such that $t+j\geq T+2N$, where $V$ is the same Lyapunov function used in the proof of Theorem \ref{main_theorem1}. We prove the claim by considering the two possible cases: (a) $t^*>0$; and (b) $t^*=0$. Suppose that case (a) holds. Then, since $t^*>0$, whenever $\tau_{v^*}(t^*,j^*)=1$, every other vertex $j$ satisfying $\tau_j(t^*,j^*)=0$ must have already reset its own state and triggered all its out-neighbors $k$. In turn, all these out-neighbors updated their state to $1$, except those who were already in $0$. However, those who are in $0$ must have already reset their own state and triggered their out-neighbors $\ell$. This argument can be repeated until all agents of the network have been exhausted, which implies that after at most $N$ jumps after the hybrid time $(t^*,j^*)$ all agents of the network have reset their state from $1$ to $0$. Finally, the same Lyapunov function used in the proof of Theorem \ref{main_theorem1} allows us to establish UGAS of $\mathcal{A}$ via the hybrid invariance principle. If case (b) holds, note that after at most $T$ seconds the root vertex $v^*$ would satisfy $\tau_{v^*}=1$, and at this point case (a) will hold. 

\emph{Necessity:} It follows from the previous observation that whenever $i^*$ is the vertex of a cycle $\mathcal{C}$, and $r=0_N$, the condition $\tau_{i^*}(0,0)=1$ will trigger sequentially all the vertices of the cycle until every vertex $k\in \mathcal{C}$ satisfies $\tau_{k}\in\{0,1\}$, with at least one vertex $j\in\mathcal{C}$ satisfying  $\tau_j=1$. By definition of cycle, the vertex $j$ will trigger at least one vertex $k$ satisfying $\tau_k=0$, generating the update $\tau_k^+=1$. This process repeats infinitely times generating a discrete solution. Therefore, if $(0_N,\mathcal{G})$ is a sync-pair, the digraph $\mathcal{G}$ cannot have cycles. \hfill $\blacksquare$ 

\begin{figure}[t!]
    \centering
    \includegraphics[width=0.11\textwidth]{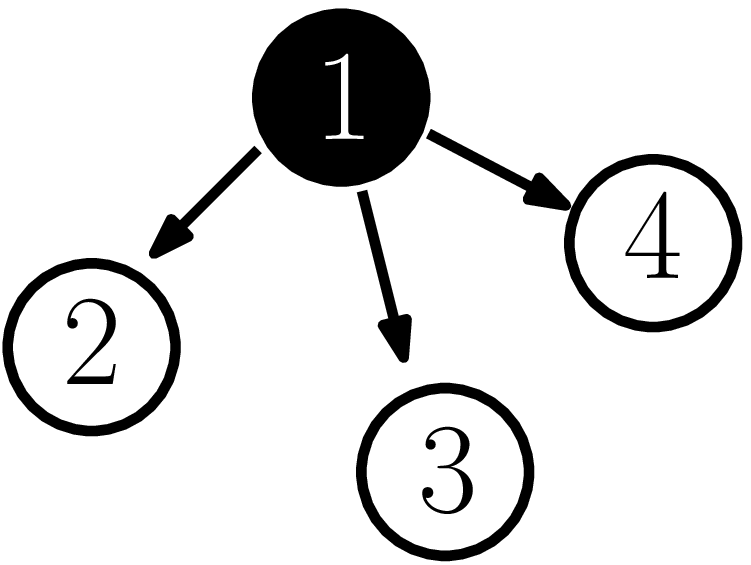}~~ \includegraphics[width=0.11\textwidth]{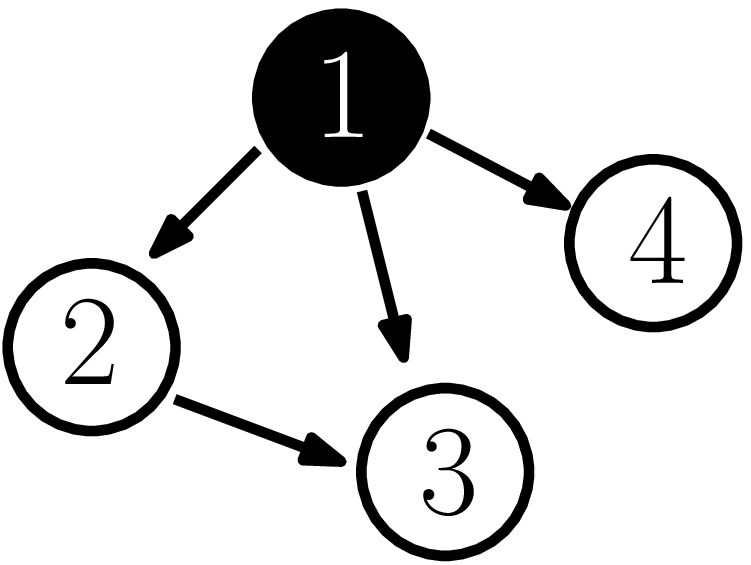}~~ \includegraphics[width=0.11\textwidth]{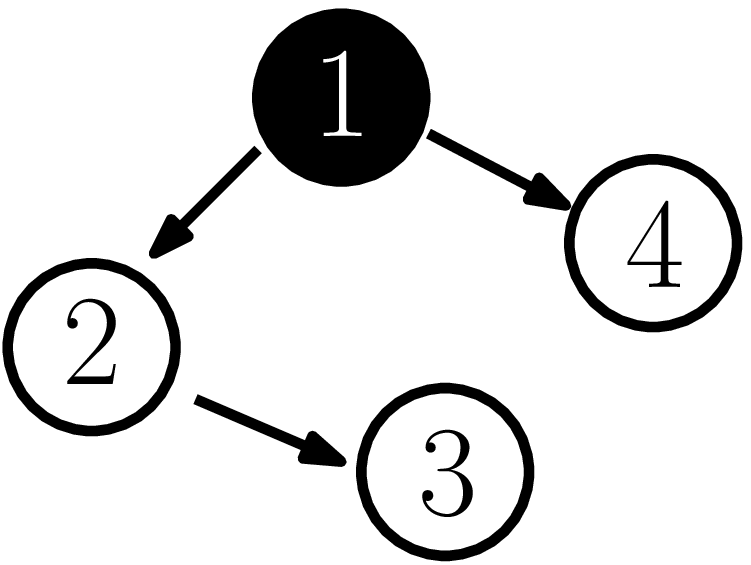}
    \caption{Rooted Acyclic Digraphs. Left: Depth=1; Center: Depth=2; Right: Depth=2. Black vertex indicates the root.}
    \label{fig:condensation}
\end{figure}

Theorems \ref{main_theorem1} and~\ref{thm:specialcaseforzeror} highlight two novel properties of PCOs with digraphs and resetting BPRs of the form \eqref{update_neigbors}: First, \emph{robust fixed-time global} synchronization can be achieved in a \emph{scalable} way for any network characterized by a rooted acyclic digraph. Indeed, in this case the bound on the parameter $r_i$ of each agent is of order $\mathcal{O}(1)$; Second, for this kind of digraphs, the synchronization can be accelerated by the parameter choice $r=0_N$, which, as noted in Remark \ref{remarkchoicer}, is prohibited if the digraphs have cycles. The results also highlight the role of the depth of the digraph in the convergence time of the hybrid dynamics. Finally, by the results of Lemma \ref{robustness_lemma}, the synchronization properties established in Theorems \ref{main_theorem1} and~\ref{thm:specialcaseforzeror} are preserved for the perturbed HDS \eqref{HDS_perturbed}, which allow us to consider small delays and drifts on the PCO's states.

\begin{remark}{\em 
While Theorems \ref{main_theorem1} and \ref{thm:specialcaseforzeror} do not cover the case where $r_i=0$ only for \emph{some} agents of the network, it is clear from the proof of Theorem \ref{thm:specialcaseforzeror} that if $\mathcal{G}$ is rooted acyclic, then $(r,\mathcal{G})$ is also a sync-pair. Therefore, there is no gap between the sufficiency results of Theorems \ref{main_theorem1} and \ref{thm:specialcaseforzeror}.}
\end{remark}

\vspace{-0.2cm}
To finish this section, we present a result that combines Lemma~\ref{lemma_1} and Theorem~\ref{main_theorem1}.   
Recall that a rooted digraph $\mathcal{G}$ is said to be quasi-acyclic if all the cycles of $\mathcal{G}$ are in its root component $\mathcal{G}_R = (\mathcal{V}_R, \mathcal{E}_R)$. The digraph $\mathcal{G}_c$ obtained by condensing the root component to a single vertex is rooted acyclic. See Fig.~\ref{fig:rootcondensation} for illustration. 

\vspace{-0.2cm}
We show that robust, fixed-time synchronization can be achieved for these digraphs as well. 
The trade-off is that the partition vector is not free to choose anymore, and there is an upper bound for every $r_i$, with $i\in \mathcal{V}_R$, of order $\mathcal{O}(\frac{1}{|\mathcal{V}_R|})$. Since the proof follows similar steps as the proof of Theorem \ref{main_theorem1}, we present the result as a corollary.
\vspace{-0.2cm}
\begin{corollary}\label{condensationcorollary}
Let $\mathcal{G}$ be rooted and quasi-acyclic, with $\mathcal{V}_R$ the root set. If $ r_i\in(0,\frac{1}{|\mathcal{V}_R| - 1})$  for any $i\in\mathcal{V}_R$, then $(r,\mathcal{G})$ is a sync-pair.  Moreover, \eqref{finite_time} holds for $T^*=(\dep(\mathcal{G}_c)+1)T$.
\end{corollary}
\vspace{-0.1cm}
\textbf{Proof:} 
By Lemma~\ref{lemma_1}, the agents in the root component will reach synchronization in no more than $T$ seconds and stay synchronized after that. We can thus treat all the roots as a whole. 
After condensing the root component to a single vertex, the resulting digraph $\mathcal{G}_c$ is rooted acyclic. Theorem~\ref{main_theorem1} then applies to the case, which completes the proof. \hfill $\blacksquare$

%

Since any strongly connected digraph $\mathcal{G}$ satisfies $\dep(\mathcal{G}_c)=0$, and any rooted acyclic digraph satisfies $\dep(\mathcal{G}_c)=\dep(\mathcal{G})$, the bound $T^*$ on the convergence time $t$ established in Corollary \ref{condensationcorollary} generalizes the bounds obtained in Lemma \ref{lemma_1} and Theorem \ref{main_theorem1}. However, as mentioned before, this generality comes at the price of the scalability of the partition vector $r$. The entries of $r$ are of order $\mathcal{O}(\frac{1}{|\mathcal{V}_R|})$, which tend to $0$ as $|\mathcal{V}_R|\to\infty$. Nevertheless, as we will show in the next section, the scalability property of $r$ can be fully recovered by adding suitable randomness into the PCOs.
\vspace{-0.2cm}
\begin{remark}\emph{
Given that Theorems \ref{main_theorem1}-\ref{thm:specialcaseforzeror}, and Corollary \ref{condensationcorollary} guarantee fixed-time synchronization of the PCOs, it is clear that all our results also hold if the digraph $\mathcal{G}_t$ is time-varying and $(T^*,L)$-\emph{persistently rooted acyclic} \cite[Def. 3]{sync_hybrid_poveda}, i.e., if for each interval $I$ of length $L$ there exists a sub-interval $I_{i}=[t_i,t_{i+1}]\subset I$ satisfying $t_{i+1}-t_i=T^*$ and a rooted acyclic digraph $\mathcal{G}^*$ such that $\mathcal{G}_t=\mathcal{G}^*$ for all $t\in I_i$.}
\end{remark}

\vspace{-0.2cm}
\section{Stochastic Resetting Algorithms}
\label{section_stochastic_sync}
\vspace{-0.2cm}
In this section, we consider networks of PCOs implementing the same hybrid update rule \eqref{flows_agents}, \eqref{resets_uncoupled}, and \eqref{update_neigbors}, but with the underlying communication network being a random digraph. In this setting, every time an agent resets its phase to~$1$, it generates {\em i.i.d.} Bernoulli random variables to decide whether or not to send pulses to its out-neighbors. In order to formalize the model of the system, we will use the framework of set-valued stochastic hybrid dynamical systems (SHDS) \cite{rec_principle}. 
\vspace{-0.2cm}
\subsection{Well-Posed Stochastic Hybrid Model}
\vspace{-0.2cm}
To formalize the model of the PCOs with random digraphs, we start by fixing a deterministic digraph $\mathcal{G}:=(\mathcal{V},\mathcal{E})$. 
Let $\mathcal{G}' = (\mathcal{V}, \mathcal{E}')$ be a subgraph of $\mathcal{G}$, with the same vertex set $\mathcal{V}$ and $\mathcal{E}' \subseteq \mathcal{E}$. 
We call any such digraph $\mathcal{G}'$ a \textbf{\textit{feasible}} digraph. 
Note that every feasible digraph $\mathcal{G}'$ can be represented by a binary vector $\psi\in \{0,1\}^{|\mathcal{E}|}$ as follows: 
\begin{equation}\label{vectorization}
v:=[\ldots, v_{ij}, \ldots],
\end{equation} 
where each entry $v_{ij}$ indicates whether $(i,j)\in \mathcal{E}$ is an edge of $\mathcal{G}'$ or not: If $v_{ij} = 1$, then $(i,j)\in \mathcal{E}'$. Otherwise, $(i,j)\not\in \mathcal{E}'$. 
Note that the binary vectors in $\{0,1\}^{|\mathcal{E}|}$ one-to-one correspond to the feasible digraphs. For convenience, we will let 
$$\Psi:= \{0,1\}^{|\mathcal{E}|},$$ 
be the set of all feasible digraphs represented by the binary vectors~$v$.  
   
\vspace{-0.2cm}
We next consider an Erd{\"o}s-R{\'e}nyi type random graph model for generating a feasible digraph. For a given vector $v\in \Psi$, we let the entries $v_{ij}$ be {\em i.i.d.} Bernoulli $(p)$ random variables, i.e.,  the probability that $v_{ij}$ takes value $1$ (resp. $0$) is $p$ (resp. $(1- p)$).  
We denote by $\mu$ the probability measure for the random graph. It follows that for any feasible digraph $\mathcal{G}'= (\mathcal{V}, \mathcal{E}')$, 
\begin{equation}\label{probabilitygraph}
\mu(\mathcal{G}') = p^{|\mathcal{E}'|} (1 - p)^{|\mathcal{E}| - |\mathcal{E}'|}.
\end{equation}

\vspace{-0.2cm}
We will now adapt the resetting algorithm to accommodate the above random graph model.  
First, note that the communication digraph affects (only) the jump map of the hybrid dynamics~\eqref{eq:DHDS}. In the previous deterministic setting, the digraph is always given by $\mathcal{G}$.  
For the stochastic setting, we replace $\mathcal{G}$ with 
a random graph $\mathcal{G}'$, with $\mathcal{G'}\sim \mu(\cdot)$. 
Furthermore, if we let $\mathcal{G}_k$, for $k \in \mathbb{N}$, be the feasible digraph at discrete time $k$ (i.e., the communication digraph at the occurrence of the $k^{th}$ jump), then all these digraphs are independent of each other. In other words, the sequence $\{\mathcal{G}_k\}^\infty_{k = 1}$ comprises {\em i.i.d.} random variables, with $\mathcal{G}_k\sim \mu(\cdot)$. 

We note here that a similar random graph model has been considered in \cite{Klinglmayr_2012,Klinglmayr}. The key difference is that the authors there considered the following scenario: Whenever an agent jumps, it draws {\em only one} Bernoulli random variable to decide whether it sends pulses to {\em all} of its out-neighbors or not. However, the digraphs considered in these two papers were either bi-directional or strongly connected. Whether their random graph model can work for rooted graphs is a non-trivial question, which we will address on another occasion.
Another major difference from the works in \cite{Klinglmayr_2012,Klinglmayr} is that we utilize tools from stochastic hybrid dynamical system to analyze the well-posedness, stability, and convergence properties of the PCOs. Indeed, given that standard PCOs are hybrid dynamical systems, the addition of randomness into the model naturally leads to a stochastic hybrid setting. 

\vspace{-0.2cm}
To construct the corresponding stochastic hybrid dynamical system (SHDS), 
it suffices to re-define the jump map. It takes three steps to do so.   
First, for each edge $(i,j)\in\mathcal{E}$, we consider the set-valued mapping $S_{ij}:[0,1]\times \Psi\rightrightarrows[0,1]$ as follows:
\vspace{-0.2cm}
\begin{equation}\label{pre_jump_map_stochastic}
S_{ij}(\tau_j,v)=v_{ij}\overline{\mathcal{P}}(\tau_j)+(1-v_{ij})\tau_{j},
\end{equation}
where $\overline{\mathcal{P}}$ is the BPR given by \eqref{update_neigbors}, and $v_{ij}$ is the entry that corresponds to the edge $(i,j)$ in $\mathcal{E}$. 
Next, using \eqref{pre_jump_map_stochastic}, we define a new set-valued mapping $G_v^0:[0,1]^N\times \Psi\rightrightarrows\mathbb{R}^N$ as follows: 
\vspace{-0.2cm}
\begin{align}\label{initial_stochastic_jump_map}
G_v^0(\tau,v):=&\Bigg\{g\in\mathbb{R}^N:g_i=0,\notag\\
& g_j\in \left\{\begin{array}{ll} S_{ij}(\tau_j,v), & (i,j)\in\mathcal{E}\\
\{\tau_j\}, &  (i,j)\notin\mathcal{E}
\end{array}\right\},~\forall~j\neq i\Bigg\},
\end{align}
which is defined to be nonempty only when $\tau_i=1$ for some $i\in\mathcal{V}$ and $\tau_j\in[0,1)$ for $j\neq i$. Finally, the jump map for the SHDS is defined as the outer-semicontinuous hull of $G_v^0$, i.e.,
\begin{equation}\label{set_valued_jump_map}
G_v(\tau,v):=\overline{G_v^0(\tau,v)}.
\end{equation}
Note that when a jump occurs and a random graph $\mathcal{G}_k$ is drawn, not every edge of $\mathcal{G}_k$ plays a role in the jump map $G_v$. Only the edges $(i,j)$ with $\tau_i = 1$ for some $i\in \mathcal{V}$, matter. Thus, the agent $i$ does not need to know the structure of the entire graph $\mathcal{G}_k$, but rather the out-going edges incident to it. In fact, these edges are completely determined by the agent through the {\em i.i.d.} Bernoulli random variables that are generated locally by the agent itself.  
The reason of including the entire graph $\mathcal{G}_k$ in the jump map $G_v$ is rather for ease of analysis.  


%
\vspace{-0.2cm}
The following lemma establishes that the jump map $G_v$ satisfies the Basic Conditions of Definition \ref{definitionbasic1}.
\vspace{-0.1cm}
\begin{lemma}\label{wellposedjumpmap}
The set-valued mapping $G_v:[0,1]^N\times \Omega \rightrightarrows[0,1]^N$ defined by \eqref{set_valued_jump_map} satisfies condition (c) of Definition \ref{definitionbasic1}.
\end{lemma}
\vspace{-0.1cm}
\textbf{Proof:} We start by considering the set-valued map $S_{ij}$ of \eqref{pre_jump_map_stochastic}. For each fixed $\tau$, the mapping $S_{ij}$ is a summation of two measurable maps. Thus, by \cite[Prop. 14.11]{Rockafellar}, the mapping $S$ is measurable with respect to $v$. Since for each $\tau\in\mathbb{R}^N$ the mapping $G^0(\tau,v)$ in \eqref{initial_stochastic_jump_map} is constructed by assigning $0$ to the $i^{th}$ component, and $S_{ij}(\tau_j,v)$ or $\tau_j$ to the other components, it follows that $v\mapsto G^0(\tau,v)$ is also measurable. Finally, measurability of the mapping $v\mapsto\text{graph}(G(\cdot,v))$ follows by the fact that $G$ is outer semicontinuous \cite[Appendix A.2.]{sta_rec}. Since by construction $G$ is locally bounded, it follows that it satisfies the basic conditions.  \hfill $\blacksquare$

\vspace{-0.2cm}
Note that the digraph $\mathcal{G}$ and the probability $p\in [0,1]$ of the Bernoulli distribution uniquely determine the  probability space $(\Omega, \mathcal{F}, \mu)$. 
Thus, the resulting SHDS depends on three parameters, namely, $p$, $r$, and $\mathcal{G}$. 
We will write the SHDS as 
\begin{equation}\label{SHDS_graph}
\mathcal{H}_S(p, r, \mathcal{G}):=(C,f,D,G_v),
\end{equation}
where the subindex $S$ indicates that the system is stochastic.
%
\vspace{-0.2cm}
\begin{remark}\label{remark_causality}{\em 
An important standing assumption of our model is the \emph{causal} dependence of the solutions on the random graphs. In particular, note that the condition $\tau_i=r_i$, or the existence of more than one agent satisfying the condition $\tau_i=1$, leads to a set $G_{v}(\tau,v)$ in \eqref{set_valued_jump_map} that has more than one element. In this case, our model will require that each particular selection $\tau^+\in G_{v}(\tau,v)$ should not be able to anticipate the next communication graph $\mathcal{G}_k$ that will be assigned to the agents at the next jump. This causality property is intrinsic to the definition of solutions to SHDS that we consider in this paper, which is presented in Appendix \ref{solutions_SHDS}. As shown in \cite{sta_rec}, the causality property is needed in order to make use of suitable Lyapunov-based arguments for the stability analysis of the system. Causality is a standard assumption in stochastic algorithms.}
\end{remark}
\vspace{-0.2cm}
As highlighted in Remarks \ref{multiplesolutions} and \ref{remark_causality}, it is important to note that in our model for each fixed $\omega\in \Omega$ the sample path $\tau_{\omega}$ generated by the SHDS \eqref{SHDS_graph} may not be unique, and the analysis of each individual solution becomes intractable as $N$ increases. This feature makes the stability analysis of the set-valued stochastic synchronization dynamics non-trivial and differs from previous results in the literature that relied on single-valued update rules \cite{Klinglmayr_2012,Klinglmayr}. 

\vspace{-0.2cm}
\subsection{Almost Sure Global Synchronization: Stability and Attractivity}

\vspace{-0.2cm}
We recall that the compact set $\mathcal{A}_s$ is defined in~\eqref{synchronization_set}, which captures all synchronized states of the network. Similar to the definition of sync-pairs for deterministic HDS, we introduce the following definition for SHDS:  

\begin{definition}\label{compatibility2}
Let $\mathcal{A}_s$ be given in~\eqref{synchronization_set}. Let $p\in [0,1]$, $r\in [0,1]^N$, and $\mathcal{G}$ be a digraph of $N$ vertices. Then,  $(p, r, \mathcal{G})$ is a \textbf{sync-triplet} if 
\vspace{-0.2cm}
\begin{enumerate}[(a)]
\item For every initial condition in $C\cup D$ there exists non-trivial random solutions almost surely, and every maximal solution of $\mathcal{H}_{S}(p,r,\mathcal{G})$ is complete and uniformly Non-Zeno almost surely;
\item The SHDS $\mathcal{H}_S(p,r,\mathcal{G})$ renders $\mathcal{A}_s$ UGASp.
\end{enumerate}
\end{definition}

%

%
%
%

%

\vspace{-0.2cm}
Note that a \emph{necessary} condition for $(p,r,\mathcal{G})$ to be a sync-triplet is that $\mathcal{G}$ is rooted. This fact can be established by using the same arguments as in the proof of Proposition \ref{lem:rooted_necessity}. However, in contrast to the deterministic setting (cf. Proposition \ref{prop:rooted_notsufficient}), we will see soon that having a rooted digraph $\mathcal{G}$ is also a \emph{sufficient} condition for $(p,r,\mathcal{G})$ to be a sync-triplet.

For ease of presentation, we let $\omega := \omega_1\omega_2\omega_3\cdots$ be a sequence of {\em i.i.d.} random variables, with each $\omega_i \sim \mu(\cdot )$ a feasible digraph. 
We denote by $\Omega$ the collection of sample paths $\omega$. 
It should be clear that for an event:
\begin{equation*}
 \Omega':= \{\omega\in \Omega \mid  \omega_1 = v_1, \cdots, \omega_k = v_k \}
\end{equation*}
with $v_i \in \Psi$ for all $i = 1,\ldots, k$, its probability is given by $\mathbb{P}( \Omega') = \prod^k_{i = 1}\mu(v_i)$. Note that a sample path $\omega$ determines the underlying digraphs for jumps at all discrete times~$k$, for $k \ge 1$. 
A solution of $\mathcal{H}_S(p, r, \mathcal{G})$ thus depends on $\omega$ and we denote it by $\pmb{\tau}_\omega$. To state the result, we further define the following random variable:

\vspace{-0.8cm}
\begin{equation}\label{firsthitting}
\pmb{T^*}(\pmb{\tau}_{\omega}):=\inf\left\{t \mid \pmb{\tau}_{\omega}(t,j)\in\mathcal{A}_s,~(t,j)\in\text{dom}(\pmb{\tau}_{\omega})\right\},
\end{equation}
which is the first hitting-time for the solution $\pmb{\tau}_{\omega}$ entering the compact set $\mathcal{A}_s$ (i.e., the instant at which the SHDS achieves synchronization for the first time). We establish below the following result:  

\begin{theorem}\label{stochastic_theorem}
If $p \in (0,1)$, $r\in (0,1]^N$, and $\mathcal{G}$ is a rooted directed graph, then $(p, r, \mathcal{G})$ is a sync-triplet. Moreover, for any initial condition $\tau_{\omega}(0,0)$, the following holds for all positive integers $n$ and all random solutions $\pmb{\tau}_{\omega}$ of the SHDS: 

\vspace{-0.5cm}
\begin{equation}\label{eq:exponentialconvergence}
    \mathbb{P}\left ( \pmb{T^*}(\pmb{\tau}_{\omega}) > nT^* \right ) \le  \rho^n,
\end{equation}

\vspace{-0.3cm}
where $T^*$ is given in \eqref{finite_time} and $\rho\in (0,1)$ is a constant given by the following: 
\vspace{-0.3cm}
\begin{equation}\label{eq:definerho}
    \rho := 1 - \left(p^{N - 1} (1 - p)^{|\mathcal{E}| - N + 1}\right)^{\dep(G) N \underline{r}^{-1}}, 
\end{equation}
\vspace{-0.2cm}
with $\underline{r}:=\min_{i\in\mathcal{V}}r_i$.
\end{theorem}
\begin{remark}{\em 
Note that by Theorem~\ref{stochastic_theorem}, the stochastic resetting algorithm is scalable because neither $p$ nor $r_i$, for $i \in \mathcal{V}$, depends on the size $N$ of the network. We also note that the result can be further generalized by allowing different agents to have heterogeneous probabilities $p_i\in (0,1)$ for the Bernoulli random variables. The same analysis we carry out below will still apply. However, for clarity of presentation, we will only establish the result for the homogeneous case where $p$ is the same for all agents.   
}
\end{remark}

We establish below Theorem~\ref{stochastic_theorem}. 
To proceed, we first establish some preliminary lemmas. We define $\mathcal{S}(\tau_0)$ as the set of all maximal random solutions of \eqref{SHDS_graph} from the initial condition $\tau_{\omega}(0,0)=\tau_0\in\mathbb{R}^N$. For each feasible initial condition, we define the following event: 
\begin{align}
    \Omega_1(\tau_{\omega}(0,0)):= &\Big\{\omega \in \Omega \mid  \forall~\tau_{\omega}\in\mathcal{S}(\tau_{\omega}(0,0)),\exists~ i^*\in \mathcal{V}_R~\notag\\
    &~~\mbox{and}~\exists~(t^*_\omega, j^*_\omega)\in\text{dom}(\tau_{\omega})~\mbox{with $t^*_\omega\le T$}\notag\\
    &~~\mbox{s.t.}~\tau_{\omega, i^*}(t^*_\omega, j^*_\omega) = 1\Big\}. 
\end{align}
We have the following fact: 


\vspace{-0.2cm}
\begin{lemma}\label{lemmagraph2}
For any $\tau_\omega(0,0)$, $\Omega_1(\tau_\omega(0,0)) = \Omega$. 
\end{lemma}

\vspace{-0.2cm}
\textbf{Proof}: The result follows directly from the fact that a root can only be influenced by another root and by the fact that $\tau$ always increases during flows. \hfill $\blacksquare$ 

\vspace{-0.2cm}
Next, we recall a fact from graph theory: 
\vspace{-0.2cm}
\begin{lemma}\label{lemmagraph3}
For a rooted digraph $\mathcal{G}$ with a root $i^*$, there exists a directed spanning tree $\mathcal{T}_{i^*}$ with $i^*$ the unique root.
\end{lemma}
\vspace{-0.2cm}
\textbf{Proof:} One can generate a desired $\mathcal{T}_{i^*}$ using the breadth-first search algorithm~\cite{graph_algos}.  \hfill $\blacksquare$



\vspace{-0.2cm}
Note that for a given root $i^*$ of $\mathcal{G}$, there may exist multiple directed spanning trees with $i^*$ the root. In the sequel, we will fix $\mathcal{T}_{i^*}$ for each root $i^*$ so that the map $i^*\mapsto \mathcal{T}_{i^*}$ is well defined. 

\vspace{-0.1cm}
Next, we let $l$ and $L$ be two positive integers. Then, for the given $l$ and $L$ and for a given root $i^*$ of $\mathcal{G}$, we define another event as follows: 
\begin{equation}\label{eq:defineOmega2}
\Omega_2(l, L, i^*):= \{\omega\in \Omega \mid \omega_k =  \mathcal{T}_{i^*}, \,\, \forall k = l + 1,\ldots,l + L\}.
\end{equation}  
We have the following fact:

\vspace{-0.2cm}
\begin{lemma}\label{lem:probability}
For any given positive integers $l$ and $L$ and for any root $i^*$ of $\mathcal{G}$,
\vspace{-0.3cm}
\begin{equation}\label{probability}
\mathbb{P}(\Omega_2(l, L, i^*)) = \left(p^{N - 1} (1 - p)^{|\mathcal{E}| - N + 1}\right)^{L}.    
\end{equation}
\end{lemma}
\vspace{-0.5cm}
\textbf{Proof:} The result follows from the fact that the random variables $\omega_k$, for $k \ge 1$, are {\em i.i.d.} and, moreover, $\mu(\mathcal{T}_{i^*}) = p^{N - 1}(1 - p)^{|\mathcal{E}| - N + 1}$ where the integer $(N - 1)$ is the number of edges of a directed spanning tree with $N$ vertices.  \hfill $\blacksquare$

With the above lemmas, we are now in a position to prove Theorem~\ref{stochastic_theorem}: 

\vspace{-0.2cm}
\textbf{Proof of Theorem \ref{stochastic_theorem}:} We first establish the fact that $(p,r, \mathcal{G})$ is a sync-triplet. Consider again the Lyapunov function $V:[0,1]^N\to\mathbb{R}_{\geq0}$ defined as the infimum of all arcs that touch all agents on the unit circle, where the points $0$ and $1$ are identified to be the same. This function is positive definite with respect to the set $\mathcal{A}_s$, it is uniformly bounded as $V(\pmb{\tau}_{\omega})\leq 1-\frac{1}{N}$ for all $\pmb{\tau}_{\omega}\in[0,1]^N$, and it does not increase during flows of the SHDS \eqref{SHDS_graph} surely, i.e, $\dot{V}(\pmb{\tau}_{\omega})\leq 0$ for all $\pmb{\tau}_{\omega}\in C$. By construction of $V$, since the number of points occupied by agents in the circle cannot increase, we also have that $V$ does not increase during jumps. Moreover, by construction of the sets $C$ and $D$, the continuity of the mapping $f$ in \eqref{flow_map_deterministic}, and Lemma \ref{wellposedjumpmap}, the SHDS satisfies the basic conditions. Also, by using the same arguments of the proof of Lemma \ref{nonzeno}, it follows that every solution of the SHDS is almost surely complete. Thus, by the stochastic hybrid invariance principle (c.f. Theorem \ref{S_rec} in the Appendix \ref{solutions_SHDS}) in order to show UGASp of the set $\mathcal{A}_s$, it suffices to show that there does not exist complete solutions of $\pmb{\tau}_{\omega}$ that remain in a non-zero level set of the Lyapunov function almost surely. 
%
%
 Equivalently, we need to show that $\mathbb{P}(\Omega_3(\tau_\omega(0,0)))<1,$ 
 where the event $\Omega_3(\tau_\omega(0,0))$ is given by

\vspace{-0.9cm}
\begin{align*}
\Omega_3(\tau_\omega(0,0)):= \Big\{&  \omega\in \Omega \mid \exists \, c > 0 \mbox{ s.t. }~\forall~\pmb{\tau}_{\omega}\in\mathcal{S}(\tau_{\omega}(0,0)), \\ 
&V(\pmb{\tau}_{\omega}(t,j)) \ge c, \,\,  
\forall~(t,j)\in \emph{\emph{\text{dom}}}(\pmb{\tau}_{\omega})\Big\}.
\end{align*}

\vspace{-0.6cm}
To establish the above fact, we will show that there exist $\rho\in (0,1)$ and a positive $T^*$ such that for any initial condition $\tau_\omega(0, 0)$, the following holds: 

\vspace{-0.6cm}
\begin{equation}\label{eq:defineOmega4}
    \mathbb{P}(\Omega_4(\tau_\omega(0,0)))> \alpha,
\end{equation}
where the event $\Omega_4(\tau_\omega(0,0))$ is given by

\vspace{-0.9cm}
\begin{align*}
    \Omega_4(\tau_\omega(0,0)):=& \Big\{\omega \in \Omega \mid~\forall~\pmb{\tau}_{\omega}\in\mathcal{S}(\tau_{\omega}(0,0))~\mbox{and}~\forall~t \ge T^*\\
    &~\mbox{s.t.}~(t,j)\in\mbox{dom}(\tau_{\omega}),~V(\pmb{\tau}_\omega(t, j)) = 0,\Big\}.
\end{align*}

\vspace{-0.8cm}
We show below that $\alpha$ and $T^*$ can be chosen to be the following values $\alpha:= 1 - \rho$, where $\rho$ is defined in~\eqref{eq:definerho}, and $T^* := (\dep(\mathcal{G}) + 1) T$.

\vspace{-0.2cm}
First, by Lemma~\ref{lemmagraph2}, for any solution $\pmb{\tau}_\omega$, there exist a hybrid time $(t^*_\omega, j^*_\omega)$, with $t^*_\omega\le T$, and a root $i^*$ of $\mathcal{G}$ such that $\pmb{\tau}_{\omega, i^*}(t^*_\omega, j^*) = 1$. 
Conditioning on the fact that $\pmb{\tau}_{\omega, i^*}(t^*_\omega, j^*_\omega) = 1$, we consider the event $\Omega_2(j^*_\omega, j^*_\omega + L, \mathcal{T}_{i^*})$, where $L := \dep(\mathcal{T}_{i^*}) N \underline{r}^{-1}$. 
Note that by Lemma~\ref{nonzeno}, for the discrete time $j$ to increase from $j^*_\omega$ to $j^*_\omega + L$, the continuous time has to increase at least $\dep(\mathcal{T}_{i^*})T$ because otherwise, there will not be as many as 
$\dep(\mathcal{T}_{i^*})N \underline{r}^{-1}$ jumps. For convenience, we let $t^{**}$ be the time that the $(j^* + L)^{th}$ jump occurs. Then, we have just shown that $t^{**} - t^* \le \dep(\mathcal{T}_{i^*}) T \le \dep(\mathcal{G})T$.
On the other hand, by definition of the event $\Omega_2(j^*, L, i^*)$ (see~\eqref{eq:defineOmega2}), the underlying digraph during this period $[t^*, t^{**}]$ is given by the directed spanning tree $\mathcal{T}_{i^*}$. Thus, by Theorem~\ref{main_theorem1}, the solution $\pmb{\tau}_\omega$ will reach synchronization at time $t^{**}$.
Note that $(t^* + t^{**}) \le (\dep(\mathcal{G}) + 1)T = T^*$. The above arguments imply that $V(\pmb{\tau}_\omega(t, j)) = 0$,   for all $t \ge T^*$. Thus, to establish~\eqref{eq:defineOmega4}, 
it now remains to show that the probability of the event $\Omega_2(j^*, L, i^*)$ is a nonzero constant, but this is given by Lemma~\ref{lem:probability} with 
$\mathbb{P}(\Omega_2(j^*, L, i^*)) = \alpha$.

\vspace{-0.2cm}
Finally, we show that~\eqref{eq:exponentialconvergence} holds. The computation in fact follows from the above argument. First, by the Bayes rule, we have that
\begin{multline*}
    \mathbb{P}\left 
(\pmb{T^*}(\pmb{\tau}_{\omega}) >  n T^*  
\right ) 
=  \mathbb{P}\left 
(\pmb{T^*}(\pmb{\tau}_{\omega}) > (n - 1) T^*  
\right )\hdots \\
~~~~~~~\hdots\times\mathbb{P}\left 
(\pmb{T^*}(\pmb{\tau}_{\omega}) > n T^*  
\, \vert \, \pmb{T^*}(\pmb{\tau}_{\omega}) > (n - 1) T^* \right ).
\end{multline*}
Because the SHDS in our case is Markovian, the conditional probability on the right hand side of the above expression can be written as $\mathbb{P}(\pmb{T^*}(\pmb{\tau'}_{\omega'}) > T^* )$,  
where $\pmb{\tau}'_{\omega'}$ is a new solution with the initial condition $\tau'_{\omega'}(0,0)$ given by $\tau'_{\omega'}(0,0) = \tau_\omega((n - 1)T^*, j)$, for some $j$ and $\omega' := \omega_{j + 1} \omega_{j + 2} \cdots$. Note that by definition of $\Omega_4(\tau'_{\omega'}(0,0))$ and~\eqref{eq:defineOmega4}, we have that 
$$
\mathbb{P}\left 
(\pmb{T^*}(\pmb{\tau'}_{\omega'}) > T^*  
\right ) = 1- \mathbb{P}(\Omega_4(\tau'_{\omega'}(0,0))) < 1- \alpha = \rho. 
$$
It then follows that 
$$
 \mathbb{P}\left 
(\pmb{T^*}(\pmb{\tau}_{\omega}) \ge  n T^*  
\right ) < \rho  \mathbb{P}\left 
(\pmb{T^*}(\pmb{\tau}_{\omega}) \ge  (n - 1) T^*  
\right ). 
$$
The above recursive formula then implies that~\eqref{eq:exponentialconvergence} holds.  \hfill $\blacksquare$ 
\vspace{-0.2cm}
\section{Numerical Studies}
\label{sec_examples}
\begin{figure}[t!]
    \centering
   \scalebox{0.2}{\subfloat{
		\centering
		\includegraphics[width=\textwidth]{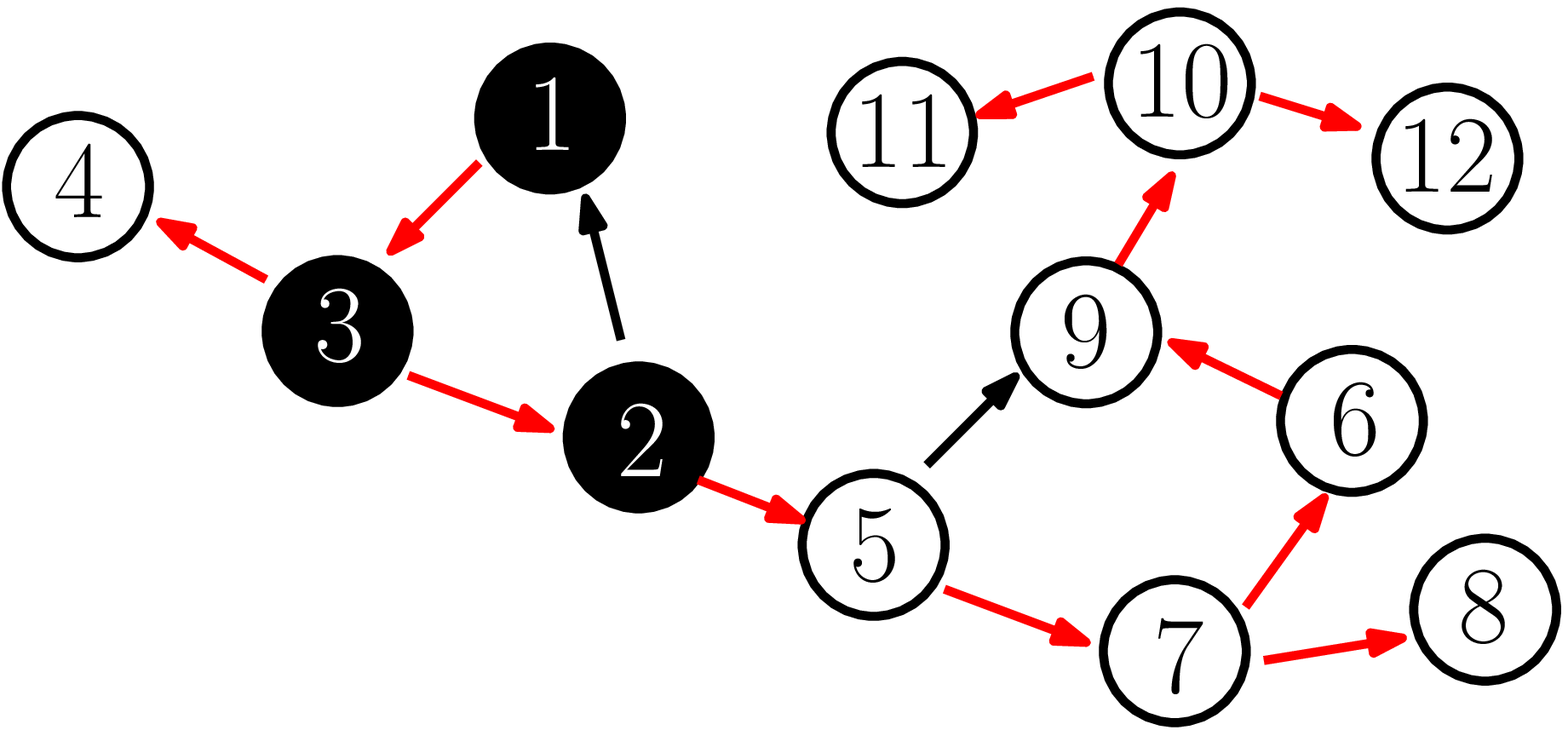}}}\hspace{1em}
	    \scalebox{0.2}{\subfloat{
			\centering
			\includegraphics[width=\textwidth]{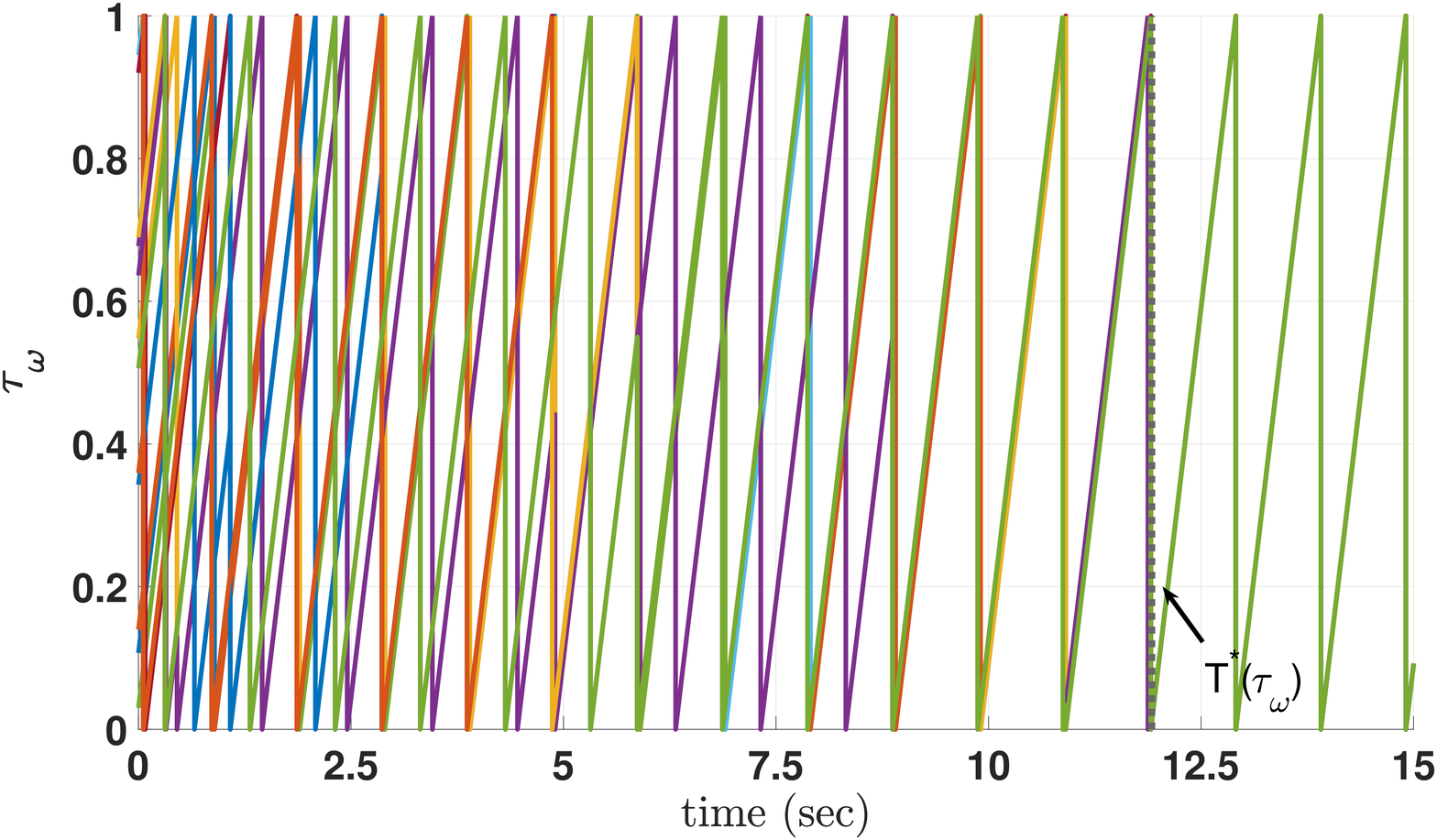}}}
    \caption{Left: Rooted digraph with nodes 1, 2, and 3 being roots. Right:  A sample path generated by the SHDS \eqref{SHDS_graph}. Synchronization is achieved in approximately 12 seconds.}
    \label{fig:simulation1}
\end{figure}

\vspace{-0.2cm}\noindent
In this section, we illustrate theoretical results by numerical examples. First, we consider a network of $N = 12$ PCOs. The underlying digraph is rooted as shown on the left of Figure \ref{fig:simulation1}. A directed spanning tree with maximum depth  is indicated by red arrows. The depth of the tree is $8$. 
We set the period of the PCOs as $T=1$, which implies that the constant $T^*$ in \eqref{finite_time}, given by $T^*:=(\dep(\mathcal{G})+1)T$, is equal to $9$. 
To simulate the SHDS \eqref{SHDS_graph}, we let the parameters $r_i$ be uniformly randomly chosen out of $(0,1)$.   
The probability $p$ of drawing an out-going edge is $0.5$.  
On the right of Figure \ref{fig:simulation1}, we show a sample path generated by the SHDS \eqref{SHDS_graph}. 

\begin{figure}[t!]
    \centering
    \includegraphics[width=0.48\textwidth]{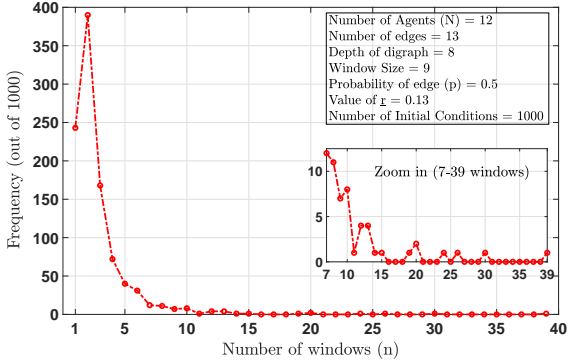}
    \caption{
    This figure shows the frequency of the number $n$ of windows needed for a sample path to achieve synchronization. There are $1000$ sample paths simulated. 
    }
    \label{fig:simulation2}
\end{figure}

Then, in Fig.~\ref{fig:simulation2}, we investigate the first hitting time $\pmb{T^*}(\pmb{\tau}_{\omega})$ defined in
 \eqref{firsthitting} using the same SHDS. We choose $1000$ random initial conditions uniformly from $(0,1)^{N}$. For each initial condition, we let $[(n-1)T^*,nT^*]$, for $n \ge 1$, be the window that contains $\pmb{T^*}(\pmb{\tau}_{\omega})$ (i.e., the sample path reaches synchronization during that period). In the figure, we plot the frequencies for different $n$. As predicted by equation \eqref{eq:exponentialconvergence}, the decay of the frequency is exponentially fast in $n$. 
 
\begin{figure}[t!]
    \centering
    \includegraphics[width=0.48\textwidth]{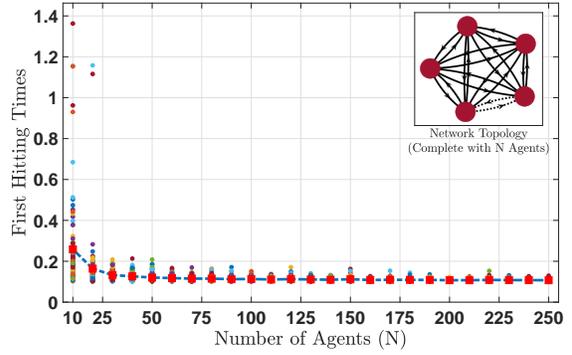}
    \caption{First hitting time vs size of complete digraph.}
    \label{fig:simulation5}
\end{figure}

\begin{figure}[t!]
    \centering
    \includegraphics[width=0.48\textwidth]{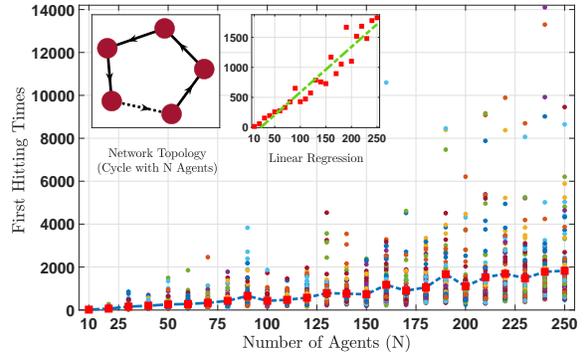}
    \caption{First hitting time vs size of cycle digraph.}
    \label{fig:simulation3}
\end{figure}

\begin{figure}[t!]
    \centering
    \includegraphics[width=0.48\textwidth]{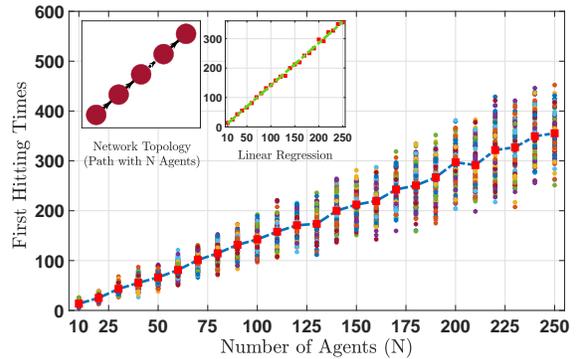}
    \caption{First hitting time vs size of path digraph.}
    \label{fig:simulation4}
\end{figure}

\vspace{-0.2cm} 
Furthermore, we investigate the dependence of the first hitting time $\pmb{T^*}(\pmb{\tau}_{\omega})$ on the size $N$ of network.  
We simulate the SHDS on three different classes of network topologies: complete digraphs, cycle digraphs, and path digraphs. 
For each class, we vary the number $N$ of agents from  $10$ to $250$, with increments of $10$.  The parameters $r_i$ are again chosen uniformly randomly from $(0,1)$ and 
the probability $p$ of drawing an edge is $0.5$.  
For each case (with a fixed class and a fixed size $N$),  we generate
 $100$ initial conditions uniformly randomly from $(0,1)^N$ and simulate the SHDS \eqref{SHDS_graph}. For each sample path, we record the first hitting time $\pmb{T^*}(\pmb{\tau}_{\omega})$. Figures \ref{fig:simulation5}, \ref{fig:simulation3}, and \ref{fig:simulation4}  plot the data for complete-, cycle-, and path-digraphs, respectively. For each figure, the horizontal axis is the network size $N$ and the vertical axis is the first hitting time $\pmb{T^*}(\pmb{\tau}_{\omega})$. For each $N$, the crosses represent the first hitting times of the sample paths. There are $100$ of them and the red square is the mean. 
 For complete digraphs, the mean decays to a steady state. The variance seems to decay as well. However, this is not the case for cycles or paths. In either case, the average and the variance increase as $N$ grows. Using linear regression, we find that the fitting curves for cycles and paths are $1.4 N - 2.6$ and $7.5N - 170.6$, respectively.


\vspace{-0.5cm}\noindent
\section{Conclusions}
\label{sec_conclusions}
\vspace{-0.2cm}\noindent
In this paper we have presented several new results in the context of synchronization of pulse-coupled oscillators evolving over sparse directed graphs. We have shown that robust global fixed-time synchronization can be achieved in quasi-acyclic digraphs using resetting algorithm, and we have characterized scalable tuning guidelines of order $\mathcal{O}(1)$ that induce this property, which can be selected to achieve acceleration in the synchronization time. We have also established an impossibility result that shows that having a rooted digraph is not sufficient for global synchronization of PCOs. However, by using suitable random communication graphs, we showed that synchronization with probability one can be established for rooted digraphs. Since all our PCOs are modeled by well-posed hybrid systems, our synchronization results are robust with respect to unmodeled dynamics and small bounded disturbances on the states. Future directions will study the development of similar distributed coordination algorithms for more general smooth compact manifolds.
\vspace{-0.2cm}
\bibliographystyle{plain}        
\bibliography{autosam}  



\appendix
\vspace{-0.2cm}
\section{Notation and Definitions}
\label{appendix_A}
\vspace{-0.2cm}
A set-valued mapping $M: \mathbb{R}^m\rightrightarrows \mathbb{R}^n$
is said to be outer semi-continuous (OSC) at $x \in \mathbb{R}^m$ if for all sequences $x_i \to x$ and $y_i \in M(x_i)$ such that $y_i \to y$ we have that $y \in M(x)$. A set-valued mapping $M: \mathbb{R}^m \rightrightarrows \mathbb{R}^n$ is said to be locally bounded (LB) at $x \in \mathbb{R}^m$ if there exists a neighborhood $K_x$ of $x$ such that $M(K_x)$ is bounded. Given a set $\mathcal{X} \subset \mathbb{R}^m$, the mapping $M$ is OSC and LB relative to $\mathcal{X}$ if the set-valued mapping from $\mathbb{R}^m$ to $\mathbb{R}^n$ defined by $M$ for $x \in \mathcal{X}$,  and by $\varnothing$ for $x \notin \mathcal{X}$, is OSC and LB at each $x \in \mathcal{X}$. The graph of a set-valued mapping $G$ is defined as $\text{graph}(G) := \{(x, y) \in  \mathbb{R}^m \times \mathbb{R}^n: y \in G(x)\}$. Given a set $B\subset\mathbb{R}^n$, we use $\text{cl}(B)$ to denote its closure.  The outer semi-continuous hull of $G$ is the unique set-valued
mapping $\overline{G}: \mathbb{R}^m \rightrightarrows\mathbb{R}^n$ satisfying $\text{graph}(\overline{G}) = \text{cl}(\text{graph}(G)$ \cite[pp. 155]{Rockafellar}. Given a measurable space $(\Omega, \mathcal{F})$, a set-valued mapping $G: \Omega \rightrightarrows \mathbb{R}^n$ is said to be \textit{$\mathcal{F}$-measurable} \cite[Def. 14.1]{Rockafellar}, if for each open set $\mathcal{O} \subset \mathbb{R}^n$, the set $G^{-1}(\mathcal{O}) :=
\{\omega \in \Omega : G(\omega) \cup \mathcal{O} 	= \varnothing\}\in \mathcal{F}$. A sequence of mappings $x_i:\text{dom}(x_i)\to\mathbb{R}^n$ is said to converge graphically if the sequence of sets $\{\text{graph}(x_i)\}_{i=1}^{\infty}$ converges in the sense of set convergence \cite[Def. 5.1]{teel}. 

\vspace{-0.3cm}
\section{Hybrid Dynamical Systems}   
\label{solutionsHDS}
\vspace{-0.2cm}
In this paper we model the dynamics of the network of pulse coupled oscillators using the formalism of hybrid dynamical systems \cite{teel}.  These systems are modeled by the equations

\vspace{-1cm}
\begin{subequations}\label{HDS1}
	\begin{align}
	&x\in C,~~~~~~~~~\dot{x}= f(x),\label{HDS_flows}\\
	&x\in D,~~~~~~x^+\in G(x),~ ~~\label{HDS_jumps}
	\end{align}
\end{subequations}
\vspace{-0.01cm}\noindent 
where $x\in\mathbb{R}^n$ is the state of the system, $f:\mathbb{R}^n\to\mathbb{R}^n$ is the flow map, which describes the continuous-time dynamics of the state; $C\subset\mathbb{R}^n$ is called the flow set and it describes the points in the space where $x$ is allowed to evolve according to the differential equation \eqref{HDS_flows}; $G:\mathbb{R}^n\times\mathbb{R}^m\rightrightarrows\mathbb{R}^n$ is the jump map and it characterizes the discrete-time dynamics of $x$; and $D\subset\mathbb{R}^n$ is called the jump set and it describes the points in the space where $x$ is allowed to evolve according to the set-valued update \eqref{HDS1}. The HDS is represented as $\mathcal{H}=\{C,D,F,G\}$. In this paper we restrict our attention to HDS that satisfy the basic conditions of Definition \ref{definitionbasic1}.
%
%
 A standard solution $x$ to \eqref{HDS1} is parameterized by a continuous-time index $t$ and a discrete-time index $j$. In particular, solutions to \eqref{HDS1} are defined on hybrid time domains. A compact hybrid time domain is a subset of $\mathbb{R}_{\geq0}\times\mathbb{Z}_{\geq0}$ of the form $\cup_{j=0}^J([t_j,t_{j+1}]\times\{j\})$ for some $J\in\mathbb{Z}_{\geq0}$ and real numbers $0=t_0\leq t_1\leq\ldots\leq t_{J+1}$. A hybrid time domain is a set $E\subset\mathbb{R}_{\geq0}\times\mathbb{Z}_{\geq0}$ such that for each $T,J$, the set $E\cap ([0,T]\times\{0,1,2,\ldots,J\})$ is a compact hybrid time domain. A function $x:E\to\mathbb{R}^n$ is said to be a hybrid arc if $E$ is a hybrid time domain, and for each $j$ such that the interval $I_j=\{t\geq0:(t,j)\in\text{dom}(x)\}$ has non-empty interior the function $t\mapsto x(t,j)$ is locally absolutely continuous. A hybrid arc $x$ is said to be a solution to a HDS \eqref{HDS1} satisfying the basic conditions if: (1) $x(0,0)\in C\cup D$. (2) If $(t_1,j), (t_2,j) \in \text{dom}(x)$ with $t_1<t_2$, then for almost every $t \in [t_1,t_2],$ $x(t,j) \in C$ and $\dot{x}(t,j) = f(x(t,j))$. (3) If $(t,j),(t,j+1) \in \text{dom}(x)$, then $x(t,j) \in D$ and $x(t,j+1) \in G(x(t,j))$.  A solution $x$ to \eqref{HDS1} is said to be: a) \emph{non-trivial} if $\text{dom}\phi$ contains at least two points; b) \emph{maximal} if there does not exist another solution $x'$ to $\mathcal{H}$ such that $\text{dom}(x)$ is a proper subset of $\text{dom}(x')$ and $x(t,j)=x'(t,j)$ for all $(t,j)\in\text{dom}(x)$; c) \emph{complete} if its domain is unbounded; d) \emph{eventually discrete} if $T=sup_t\text{dom}(x)<\infty$ and $\text{dom}(x)\cap (\{T\}\times\mathbb{N})$ contains at least two points. d) \emph{uniformly non-Zeno} if there exists $(T,J)\in\mathbb{R}_{>0}$ such that for every $(t_1,j_1),(t_2,j_2)\in\text{dom}(x)$, if $t_2-t-1\leq T$ then $j_2-j_1\leq J$.

\vspace{-0.2cm}
\section{Stochastic Hybrid Dynamical Systems}
\label{solutions_SHDS}
\vspace{-0.2cm}
When the jump map in \eqref{HDS_jumps} also depends on a random input $\bf{v}$, the HDS \eqref{HDS1} becomes a stochastic hybrid dynamical system (SHDS) \cite{rec_principle} of the form
\vspace{-0.2cm}
\begin{subequations}\label{SHDS}
	\begin{align}
	&x\in C,~~~~~~~~~\dot{x}= f(x),\label{SHDS_flows}\\
	&x\in D,~~~~~~x^+\in G(x,v^+),~~~v\sim\mu(\cdot), ~~\label{SHDS_jumps}
	\end{align}
\end{subequations}

\vspace{-0.6cm}
where $v^+$ is a place holder for a sequence $\{\bf{v}_k\}_{k=1}^{\infty}$ of independent, identically distributed {\em i.i.d.} input random variables $\bf{v}_k:\Omega\to\mathbb{R}^m$, $k\in\mathbb{N}$, defined on a probability space $(\Omega,\mathcal{F},\mathbb{P})$. Thus, ${\bf v_k}^{-1}(F):=\{\omega\in\Omega: \mathbf{v}_k(\omega)\in F\}\in\mathcal{F}$ for all $F\in\mathbf{B}(\mathbb{R})^m$, and $\mu:\mathbf{B}(\mathbb{R}^m)\to[0,1]$ is defined as $\mu(F):=\mathbb{P}\{\omega\in\Omega:{\bf v}_k(\omega)\in F\}$. We restrict our attention to SHDS that satisfy the basic conditions of Definition \ref{definitionbasic1}.
%
%
Random solutions to SHDS \eqref{SHDS} are functions of $\omega\in\Omega$ denoted ${\bf x}(\omega)$, such that: 1) $\omega\mapsto {\bf x}(\omega)$ has measurability properties that are adapted to the minimal filtration of ${\bf v}$; 2) for each $\omega\in\Omega$ the sample path ${\bf x(\omega)}$ is a standard solution to the HDS \eqref{HDS1} with the appropriate dependence on the random input ${\bf v(\omega)}$ through the jumps. To formally define these mappings, for $k\in\mathbb{Z}_{\geq1}$, let $\mathcal{F}_k$ denote the collection of sets $\{\omega\in\Omega:({\bf v}_1(\omega),{\bf v}_2(\omega),\ldots,{\bf {v}}_k(\omega))\in F\}$, $F\in\mathbf{B}(\mathbb{R}^m)^k)$, which are the sub-$\sigma$-fields of $\mathcal{F}$ that form the minimal filtration of ${\bf v}=\{{\bf v}_k \}_{k=1}^{\infty}$, which is the smallest $\sigma$-algebra on $(\Omega,\mathcal{F})$ that contains the pre-images of $\mathbf{B}(\mathbb{R}^m)$-measurable subsets on $\mathbb{R}^m$ for times up to $k$. A stochastic hybrid arc is a mapping ${\bf x}$ from $\Omega$ to the set of hybrid arcs, such that the set-valued mapping from $\Omega$ to $\mathbb{R}^{n+2}$, given by  $\omega\mapsto \text{graph}({\bf x}(\omega)):=\big\{(t,j,z):\tilde{x}={\bf x}(\omega), (t,j)\in\text{dom}(\tilde{x}),z=\tilde{x}(t,j)\big\}$, is $\mathcal{F}$-measurable with closed-values. Let $\text{graph}({\bf x}(\omega))_{\leq k}:=\text{graph}({\bf x} (\omega))\cap (\mathbb{R}_{\geq0}\times\{0,1,\ldots,k\}\times\mathbb{R}^n)$. An $\{\mathcal{F}_k\}_{k=0}^{\infty}$ adapted stochastic hybrid arc is a stochastic hybrid arc ${\bf x}$ such that the mapping $\omega\mapsto \text{graph}({\bf x}(\omega))_{\leq k}$ is $\mathcal{F}_k$ measurable for each $k\in\mathbb{N}$. An adapted stochastic hybrid arc ${\bf x}$ is a solution to \eqref{SHDS} starting from $x_0$ denoted ${\bf x}\in \mathcal{S}_r(x_0)$ if (with $x_{\omega}:={\bf x}(\omega)$): (1) $x_{\omega}(0,0)=x_0$; (2) if $(t_1,j),(t_2,j)\in\text{dom}(x_{\omega})$ with $t_1<t_2$, then for all $t\in[t_1,t_2]$, $x_{\omega}(t,j)\in C$ and $\dot{x}_{\omega}(t,j)= f(x_{\omega}(t,j))$; (3) if $(t,j),(t,j+1)\in\text{dom}(x_{\omega})$, then $x_{\omega}(t,j)\in D$ and $x_{\omega}(t,j+1)\in G(x_{\omega}(t,j),\mathbf{v}_{j+1}(\omega))$. A random solution ${\bf x}$ is said to be: a) almost surely complete if for almost every sample path $\omega\in \Omega$ the hybrid arc ${\bf x}(\omega)$ has an unbounded time domain; and almost surely eventually discrete if for almost every sample path $\omega\in \Omega$ the hybrid arc ${\bf x}(\omega)$ is eventually discrete.

A continuous function $V:\mathbb{R}^n\to\mathbb{R}_{\geq0}$ is a Lyapunov function relative to a compact set $\mathcal{A}\subset\mathbb{R}^n$ for the SHDS \eqref{SHDS} if $V(x)=0\iff x\in\mathcal{A}$, $V$ is radially unbounded and satisfies $V(\phi(t))\leq V(x),~~\forall~x\in C,~t\in\text{dom}(\phi),~\phi\in S^F_C(x)$, and $\int_{R^m}\max_{g\in G(x,v)} V(g)\mu(dv)\leq V(x),~~\forall~x\in D$, where $S^F_C(x)$ denotes the set of solutions of \eqref{SHDS_flows} with initial condition $x$. The following stochastic hybrid invariance principle, corresponding to \cite[Thm. 8]{rec_principle}, is instrumental in the analysis of SHDS of the form \eqref{SHDS}.
\begin{thm}\label{S_rec}
Let $V$ be a Lyapunov function relative to a compact set $\mathcal{A} \subset \mathbb{R}^n$ for the SHDS system $\mathcal{H}$. Then, $\mathcal{A}$  is UGASp if and only if there does not exist an almost surely complete solution $x$ that remains in a non-zero level set of the Lyapunov function almost surely. 
\end{thm}

\noindent \textbf{Umar Javed} received in 2017 his B.S. degree in Electrical Engineering with minors in Computer Science (AI/ML) and Psychology from the Lahore University of Management Sciences, Pakistan. Currently, he is a PhD candidate in the Department of Electrical, Computer and Energy Engineering at the University of Colorado, Boulder, where he completed his MS in Electrical Engineering in 2019. His interests lie in mathematical control theory and its applications

\noindent \textbf{Jorge I. Poveda} is an Assistant Professor in the Department of Electrical, Computer, and Energy Engineering at the University of Colorado, Boulder. He received the M.Sc. and Ph.D. degrees in Electrical and Computer Engineering from the University of California at Santa Barbara in 2016 and 2018, respectively, where he was awarded the Center for Control Dynamical Systems and Computation Outstanding Scholar Fellowship. Before joining CU Boulder, he was a Postdoctoral Fellow at Harvard University in 2018, and a research intern at the Mitsubishi Electric Research Laboratories in 2016 and 2017. In 2020 he received the NSF Career Research Initiation Award (CRII). His research interests lie in the design and analysis of high-performance feedback-based control and optimization algorithms for cyber-physical systems.           
                                  
\noindent \textbf{Xudong Chen} is an Assistant Professor in the Department of Electrical, Computer and Energy Engineering at the University of Colorado, Boulder. Before that, he was a postdoctoral fellow in the Coordinated Science Laboratory at the University of Illinois, Urbana-Champaign. He obtained the B.S. degree from Tsinghua University, Beijing, China, in 2009, and the Ph.D. degree in Electrical Engineering from Harvard University, Cambridge, Massachusetts, in 2014, under the supervision of Roger Brockett. His research interests are in the area of control theory, stochastic processes, optimization, game theory and their applications in modeling, control, and estimation of networked systems and ensemble systems. His research group develops novel engineering methods and advanced mathematical tools for investigating large-scale multi-agent systems. Dr. Chen is an awardee of the 2020 Air Force's Young Investigator Research Program (YIP).                                     

\end{document}